\documentclass[aps,10pt,pra,twocolumn,showpacs,showkeys,groupedaddress,nofootinbib]{revtex4-1}
\usepackage[colorlinks,linkcolor=red,citecolor=blue,urlcolor=blue]{hyperref}
\usepackage{array}
\usepackage{amsmath}
\usepackage{graphicx}
\usepackage{bm}
\usepackage{xcolor}
\usepackage{comment}
\usepackage[normalem]{ulem}
\usepackage{enumitem}
\usepackage{natbib}
\usepackage{textcomp}
\usepackage{microtype}
\usepackage{multirow}

\newcommand{\be}{\begin{equation}}
\newcommand{\ee}{\end{equation}}
\newcommand{\bea}{\begin{eqnarray}}
\newcommand{\eea}{\end{eqnarray}}

\def\mum{\rm\mu m}
\def\mus{\rm\mu s}

\def\ie{{\it i.e.,\/}}
\def\eg{{\it e.g.,\/}}
\def\vs{{\it vs.\/}}

\def\etc{{\it etc.\/}}

\def\etal{{\it et al.\/}}

\def\Rb{$\rm^{87}Rb$}

\def\one{|1\rangle}
\def\two{|2\rangle}

\def\halfskip{\vskip0.5\baselineskip}

\def\negskip {\vskip  -\baselineskip}
\setlist[enumerate]{wide=-4pt,nolistsep}
\begin{document}

\title{Stern-Gerlach Interferometry with the Atom Chip}

\author{Mark~Keil}
	\thanks{\footnotesize Corresponding authors:  
	\href{mailto:mhkeil@gmail.com}{\tt mhkeil@gmail.com}, \href{mailto:folman@bgu.ac.il}{\tt folman@bgu.ac.il}}
	\affiliation{Department of Physics, Ben-Gurion University of the Negev, Be'er Sheva 84105, Israel}
\author{Shimon~Machluf}	
	\thanks{\footnotesize Present address: Analytics Lab, Amsterdam, The Netherlands}
	\affiliation{Department of Physics, Ben-Gurion University of the Negev, Be'er Sheva 84105, Israel}
\author{Yair~Margalit}	
	\thanks{\footnotesize Present address: Research Laboratory of Electronics, MIT-Harvard Center for Ultracold Atoms, Department of Physics, Massachusetts Institute of Technology, Cambridge, MA 02139, USA}
	\affiliation{Department of Physics, Ben-Gurion University of the Negev, Be'er Sheva 84105, Israel}
\author{Zhifan~Zhou}	
	\thanks{\footnotesize Present address: Joint Quantum Institute, National Institute of Standards and Technology and the University of Maryland, College Park, Maryland 20742 USA}
	\affiliation{Department of Physics, Ben-Gurion University of the Negev, Be'er Sheva 84105, Israel}
\author{Omer~Amit}
	\affiliation{Department of Physics, Ben-Gurion University of the Negev, Be'er Sheva 84105, Israel}
\author{Or~Dobkowski}	
	\affiliation{Department of Physics, Ben-Gurion University of the Negev, Be'er Sheva 84105, Israel}
\author{Yonathan~Japha}	
	\affiliation{Department of Physics, Ben-Gurion University of the Negev, Be'er Sheva 84105, Israel}
\author{Samuel~Moukouri}	
	\affiliation{Department of Physics, Ben-Gurion University of the Negev, Be'er Sheva 84105, Israel}
\author{Daniel~Rohrlich}	
	\affiliation{Department of Physics, Ben-Gurion University of the Negev, Be'er Sheva 84105, Israel}
\author{Zina~Binstock}	
	\affiliation{Department of Physics, Ben-Gurion University of the Negev, Be'er Sheva 84105, Israel}
\author{Yaniv~Bar-Haim}	
	\affiliation{Department of Physics, Ben-Gurion University of the Negev, Be'er Sheva 84105, Israel}		
\author{Menachem~Givon}	
	\affiliation{Department of Physics, Ben-Gurion University of the Negev, Be'er Sheva 84105, Israel}	
\author{David~Groswasser}	
	\affiliation{Department of Physics, Ben-Gurion University of the Negev, Be'er Sheva 84105, Israel}
\author{Yigal~Meir}	
	\affiliation{Department of Physics, Ben-Gurion University of the Negev, Be'er Sheva 84105, Israel}
\author{Ron~Folman}
	\thanks{\footnotesize Corresponding authors:  
	\href{mailto:mhkeil@gmail.com}{\tt mhkeil@gmail.com}, \href{mailto:folman@bgu.ac.il}{\tt folman@bgu.ac.il}}
	\affiliation{Department of Physics, Ben-Gurion University of the Negev, Be'er Sheva 84105, Israel}	
	
\makeatletter
\renewcommand\frontmatter@abstractwidth{\dimexpr\textwidth-24mm\relax}
\makeatother

\begin{abstract}

{\parindent=0pt 

In this invited review in honor of~100 years since the Stern-Gerlach~(SG) experiments, we describe a decade of~SG interferometry on the atom chip. The~SG effect has been a paradigm of quantum mechanics throughout the last century, but there has been surprisingly little evidence that the original scheme, with freely propagating atoms exposed to gradients from macroscopic magnets, is a fully coherent quantum process. Specifically, no full-loop~SG interferometer~(SGI) has been realized with the scheme as envisioned decades ago. Furthermore, several theoretical studies have explained why it is a formidable challenge. Here we provide a review of our~SG experiments over the last decade. We describe several novel configurations such as that giving rise to the first~SG spatial interference fringes, and the first full-loop~SGI realization. These devices are based on highly accurate magnetic fields, originating from an atom chip, that ensure coherent operation within strict constraints described by previous theoretical analyses. Achieving this high level of control over magnetic gradients is expected to facilitate technological applications such as probing of surfaces and currents, as well as metrology. Fundamental applications include the probing of the foundations of quantum theory, gravity, and the interface of quantum mechanics and gravity. We end with an outlook describing possible future experiments.} 

\end{abstract}

\date{\today}

\maketitle

\twocolumngrid
\parindent=0pt
\parskip=0.75\baselineskip
\negskip

\section{Introduction}\negskip
\label{sec:introduction}

This review follows the centennial conference held in Frankfurt in the same building housing the original Stern-Gerlach~(SG) experiments. Here we describe the~SG interferometry performed in our laboratories at Ben-Gurion University of the Negev~(BGU) over the last decade.

The trail-blazing experiments of Otto Stern and Walther Gerlach one hundred years ago~\cite{Gerlach1922,Friedrich2019,Friedrich2020,Herschbach2001}  required a few basic ingredients: a source of isolated atoms with well-specified momentum components (provided by their atomic beam), an inhomogeneous magnetic field and, if we follow the historical account of events in~\cite{Friedrich2003}, also a smoky cigar. In this review, we present our approach to these first two ingredients, with our sincere apologies that we will not be able to adequately address the third.

As Dudley Herschbach notes~\cite{Herschbach2001}, the~SG experiments formed the basis for a ``symbiotic entwining of molecular beams with quantum theory'' and, as shown in many of the papers at this centennial conference, this symbiotic relationship remains vigorous to the present day. In this review, our source of isolated atoms is instead provided by the new world of ultra-cold atomic physics, to which we couple inhomogeneous magnetic fields that are provided naturally by an atom chip~\cite{Keil2016}. Current-carrying wires on such chips were first realized as magnetic traps for ultra-cold atoms at the turn of the (21$^{\rm st}$) century~\cite{Reichel1999,Folman2000,Dekker2000} and reviewed extensively since~\cite{Folman2002,Reichel2002,Fortagh2007,Reichel2011_book,Folman2011a,Keil2016}. We are using the atom chip as our basis for coherently manipulating atoms in a way that is complementary to the atomic and molecular beam techniques pioneered by Otto Stern and practiced so energetically and creatively by his scientific descendants. 

The work presented here is performed with high-quality atom chips fabricated by our nano-fabrication facility~\cite{nanofab}. The atom chip is advantageous for Stern-Gerlach interferometry~(SGI) for~4 main reasons. First, the source~(Bose-Einstein condensates,~BEC) is a minimal-uncertainty wavepacket so it is very well defined in position and momentum. Second, the source of the magnetic gradients (current-carrying wires on the atom chip) is very well aligned relative to the atomic source. Third, due to the very small atom-chip distance, the gradients are very strong, and significant Stern-Gerlach splitting can be realized in very short times. Fourth, the gradients are very well defined in time since there are no coils and the inductance of the chip wires is negligible. We will describe how these advantages have overcome long-standing difficulties and have enabled different~SG configurations to be realized at~BGU (\eg\ spatial interference patterns~\cite{Machluf2013,Margalit2019} and a ``full-loop''~SGI~\cite{Margalit2018x,Amit2019}) alongside several applications, such as spatially splitting a clock~\cite{Margalit2015,Zhou2018}. Finally, let us mention that while the  interferometers presented here are of a new type, it is worthwhile noting decades of progress in matter-wave interferometry~\cite{Cronin2009}.

The discovery of the Stern-Gerlach~(SG) effect~\cite{Gerlach1922} was followed by ideas concerning a full-loop~SGI that would consist of freely propagating atoms exposed to magnetic gradients from macroscopic magnets. However, starting with Heisenberg~\cite{Heisenberg1930}, Bohm~\cite{Bohm1951} and Wigner~\cite{Wigner1963} considered a coherent~SGI impractical because it was thought that the macroscopic device could not be made accurate enough to ensure a reversible splitting process~\cite{Briegel1997}. Bohm, for example, noted that the magnet would need to have ``fantastic'' accuracy~\cite{Bohm1951}. Englert, Schwinger and Scully analyzed the problem in more detail and coined it the Humpty-Dumpty\footnote{
Can a fragile item be taken apart and be re-assembled perfectly?\\
\dots~another tough problem, according to the popular English rhyme~\cite{HDrhyme}\\
Humpty Dumpty sat on a wall,\\
Humpty Dumpty had a great fall.\\
All the king's horses\\ 
And all the king's men\\
Couldn't put Humpty together again.}~(HD)
effect~\cite{Englert1988,Schwinger1988,Scully1989,Englert1997}. They too concluded that for significant coherence to be observed, exceptional accuracy in controlling magnetic fields would be required. Indeed, while atom interferometers based on light beam-splitters enjoy the quantum accuracy of the photon momentum transfer, the~SGI magnets not only have no such quantum discreteness, but they also suffer from inherent lack of flatness due to Maxwell's equations~\cite{Scully1987}. Later work added the effect of dissipation and suggested that only low-temperature magnetic field sources would enable an operational~SGI~\cite{Oliveira2006}. Claims have even been made that no coherent splitting is possible at all~\cite{Devereux2015}.

Undeterred, we utilize the novel capabilities of the atom chip to address these significant hurdles. Let us briefly preview our most recent and most challenging realization, the full-loop~SGI, in which magnetic field gradients act on the atom during its flight through the interferometer, first splitting, and then re-combining, the atomic wavepacket. We obtain a high full-loop~SGI visibility of~95\% with a spin interference signal~\cite{Margalit2018x,Amit2019} by utilizing the highly accurate magnetic fields of an atom chip~\cite{Keil2016}. Notwithstanding the impressive endeavors of~\cite{Robert1991,Miniatura1991,Miniatura1992a,Robert1992,Miniatura1992b,Chormaic1993,Baudon1999,Boustimi2000,Marechal2000,Lesegno2003,Rubin2004} this is, to the best of our knowledge, the first realization of a complete~SG interferometer analogous to that originally envisioned a century ago.
 
Achieving this high level of control over magnetic gradients may facilitate fundamental research. Stern-Gerlach interferometry with mesoscopic objects has been suggested as a compact detector for space-time metric and curvature~\cite{Marshman2020b}, possibly enabling detection of gravitational waves. It has also been suggested as a probe for the quantum nature of gravity~\cite{Bose2017}. Such~SG capabilities may also enable searches for exotic effects like the fifth force or the hypothesized self-gravitation interaction~\cite{Hatif2020x}. We note that the realization presented here has already enabled the construction of a unique matter-wave interferometer whose phase scales with the cube of the time the atom spends in the interferometer~\cite{Amit2019}, a configuration that has been suggested as an experimental test for Einstein's equivalence principle when extended to the quantum domain~\cite{Marletto2020}.
 
High magnetic stability and accuracy may also benefit technological applications such as large-momentum-transfer beam splitting for metrology with atom interferometry~\cite{Gebbe2019x,Canuel2020x,Rudolph2020}, sensitive probing of electron transport, \eg~squeezed currents~\cite{Strekalov2011}, as well as nuclear magnetic resonance and compact accelerators~\cite{Danieli2013}. We note that since the~SGI makes no use of light, it may serve as a high-precision surface probe at short distances for which administering light is difficult.

For the purpose of this review, it is especially important to also realize that the atom chip allows our atoms to be completely isolated from their environment. This is demonstrated, for example, by the relatively long-term maintenance of spatial coherence that can be achieved despite a temperature gradient from~$\rm300\,K$ to~$\rm100\,nK$ over a distance of just~$\rm5\,\mum$~\cite{Zhou2016}. Coherence of internal degrees of freedom close to the surface has also been measured to be very high~\cite{Treutlein2004}. 

This review is organized into the following sections:\\[-12pt]
\vspace{-\topsep}
\begin{itemize}[noitemsep,leftmargin=0pt]
\item[]\ref{sec:sources}. Particle Sources: a brief discussion of how the atom chip complements and extends the century-long use of atomic and molecular beams in Stern-Gerlach experiments;\\[-6pt]
\item[]\ref{sec:SGBS}. The Atom Chip Stern-Gerlach Beam Splitter: detailing relevant aspects of the atom chip and its basic operating characteristics as a platform for~SGI;\\[-6pt]
\item[]\ref{sec:halfloop}. Half-Loop Stern-Gerlach Interferometer: first realization of~SGI with spatial fringe patterns;\\[-6pt]
\item[]\ref{sec:fullloop}. Full-Loop Stern-Gerlach Interferometer: first realization of the four-field complete SGI with spin population fringes;\\[-6pt] 
\item[]\ref{sec:applications}. Applications: clock interferometry and complementarity, the matter-wave geodesic rule and geometric phase, and a~$T^3$ interferometer realizing the Kennard phase;\\[-6pt]
\item[]\ref{sec:outlook}. Outlook: extending the atom-chip based~SGI experiments to ion beams and to massive particles.
\end{itemize}
\vspace{-\topsep}

Finally, we note that the~SG effect, in conjunction with the atom chip, may also be used for novel applications without the use of interferometry. For example, we have used the~SG spin-momentum entanglement to realize a novel quantum work meter. In this work, done in conjunction with the group of Juan Pablo Paz, we were able to test non-equilibrium fluctuation theorems~\cite{Cerisola2017}.

As we hope to show in this review, we believe that the atom chip provides a novel and powerful tool for~SG interferometry, with much yet to learn as~SG studies enter their second century. May we continue to find surprises, fundamental insights, and exciting applications.

{\parindent=0pt
{
\renewcommand{\arraystretch}{1.25}
	\begin{table*} 
  \centering	
\begin{tabular*}{1.00\textwidth}{@{\extracolsep{\fill}}l l c c c c c c c}\hline
  \multirow{2}{*}{type}
& \multirow{2}{*}{source}  
& \multirow{2}{*}{species}
& temperature  
& $\sigma_z$  
& $\sigma_{v_z}$  
& $k=\sigma_{p_z}/\hbar$  
& \multirow{2}{*}{$\sigma_z\sigma_{p_z}/\hbar$}	
& \multirow{2}{*}{Ref.}	                      
\\[-3pt]
                       &          &                & (K)          & ($\mum$) & (mm/s) & ($\mum^{-1}$) &				   &                             \\\hline
diffraction            & beam	    & $^{4}$He	     & $10^{-3}$	  & 20			 & 14	    & 0.9	          & 18	     & \cite{Zhao2013}	           \\ 
diffraction            & beam	    & $^{4}$He	     & not given	  & 50			 & 43	    & 2.6	          & 130      & \cite{Zeller2016}           \\ 
T-L interference$^1$   & beam		  & macromolecules & not given	  & 0.266		 & 0.04   & 16	          & 4.3      & \cite{Fein2019}             \\ 		
interference           & BEC		 	& \Rb				     & 40x$10^{-9}$	& 6  		   & 2.8	  & 3.8	          & 23	     & \cite{Margalit2019}         \\ 
particle-on-demand$^2$ & ion trap & $^{40}$Ca$^+$  &	--	    	  & 0.006    & 900    & 5x10$^6$      & 3x10$^4$ & \cite{Jacob2016,Henkel2019} \\ 
first realization      & beam     & Ag             & 1300         & 30       & 230    & 400           & 1x10$^4$ & \cite{Gerlach1922}          \\\hline
\end{tabular*}
\parbox{1.00\textwidth}{\caption{\label{table:beams} Parameters relating to diffraction experiments using~He atomic beams~\cite{Zhao2013,Zeller2016},
Talbot-Lau interference experiments with macromolecules~\cite{Fein2019}, and interference experiments using BEC's~\cite{Margalit2019} as described in this review. The temperature shown for the beam experiments corresponds to the velocity spread superimposed on the moving frame of the longitudinal most-probable velocity. The position spread~$\sigma_z$ for the~He beams is the beam collimator width, while the velocity spread~$\sigma_{v_z}$ is calculated from the beam angular divergence and its most-probable longitudinal velocity~\cite{Bruch2002} ($v_x=v_{\rm mp}\approx\rm288\,m/s$ for a~He beam source temperature of~$\rm8\,K$). For the macromolecular beam, the parameters are taken from the grating period, interferometer length, and the stated longitudinal deBroglie wavelength. Corresponding parameters for the~BEC are calculated using the Thomas-Fermi approximation and a temperature at which the~BEC is about~90\% pure. Parameters for the original Stern-Gerlach experiment are shown for comparison in the last line. All species are in their ground electronic state. The~$x$- and~$z$- co-ordinates refer to the horizontal and vertical directions respectively, where the beam experiments are horizontal (so~$z$ is the transverse direction) while the~BEC experiments are vertical (so~$z$ is the longitudinal direction). We do not give parameters for the ``beaded atom'' experiments~\cite{Miniatura1991} since we believe that spatial interference fringes were not observed, as explained in~\cite{Margalit2018t}.
\\$^1$ Talbot-Lau interference, as applied to matter-wave interference studies, is described in detail in~\cite{Kialka2019}. The particle species in the quoted study are functionalized oligoporphyrin macromolecules with up to~2000 atoms and masses~$>\rm25000\,Da$~\cite{Fein2019}.
\\$^2$ These parameters are for a proposal for~SGI using ion beams that will be discussed in Sec.~\ref{subsec:ionSGI}.}}
\end{table*}}
}

\negskip\negskip
\section{Particle Sources}\negskip
\label{sec:sources}

Molecular beam experiments exhibiting quantum interference, diffraction, and reflection have been brought very skillfully into the modern era in presentations at this Conference by Markus Arndt, Maksim Kunitski, and Wieland Sch\"ollkopf, and as outlined in the keynote address by Peter Toennies. In particular, Stern's vision~--~and realization~--~of diffraction of atomic and molecular beams (see, for example~\cite{Herschbach2001}) have found their modern expression in the work of all these experts, and many others. Here we will concentrate on a complementary approach to precisely specify internal and external quantum states and how they can be used to study interference phenomena in particular. 

Let us begin by comparing experimental parameters used in the ultra-cold atomic environment in our laboratory, typically achieved with~BECs of~\Rb, with corresponding state-of-the-art parameters for atomic beams. Table~\ref{table:beams} summarizes parameters that are most relevant for these experiments. Note that the beam experiments are conducted in a horizontal plane, transverse to the beam propagation direction, while our~BEC interference experiments are conducted in an exclusively longitudinal direction with the atoms falling vertically due to gravity (and with all applied forces also acting in the longitudinal direction).  

We see that ultra-cold atom localization and velocity spreads are on the same order as transverse localization from the exemplary atomic and molecular beam experiments quoted here but, of course, ultra-cold atoms are also localized in all three dimensions, whereas the beam techniques do not achieve localization along the beam propagation axis.

\negskip\negskip
\section{The Atom Chip Stern-Gerlach Beam Splitter}\negskip
\label{sec:SGBS}

In order to apply Stern-Gerlach splitting, our ultra-cold atomic sample needs to have at least two spin states. However, our initial atomic sample is purely in the~$|F,m_F\rangle=|2,2\rangle$ state of~\Rb. After preparing a~BEC on the atom chip, our~SG implementation therefore begins by first releasing the magnetic trap, and then applying a radio-frequency~(RF)~$\pi/2$ Rabi pulse to create an equal superposition of the two internal spin states~$\frac{1}{\sqrt2}(\one+\two)$, where~$\one$ and~$\two$ represent the~$m_F=1$ and~$m_F=2$ Zeeman sub-levels of the~$F=2$ manifold in the ground electronic state~\cite{Steck2003}. Transitions to other~$m_F$ levels are avoided by retaining a modest homogeneous magnetic field even after trap release. A field of about~$30\rm\,G$ is sufficient to create an effective two-level system by pushing the~$m_F=0$ sub-level about~$\rm200\,kHz$ out of resonance with the~$\two\to\one$ RF transition due to the non-linear Zeeman effect. The intensity of the~RF Rabi pulses is calibrated such that a pulse duration of~$\rm20\,\mus$ corresponds to a complete population inversion between the two states, \ie\ a~$\pi$-pulse. This corresponds to a Rabi frequency of~$\Omega_{\rm RF}=\rm2\pi\cdot25\,kHz$.

We now consider the second factor crucial to the success of our~SGI experiments: fast and precise magnetic fields, in both magnitude and direction, may be delivered by pulsed currents passed through micro-fabricated wires on the atom chip. Simple Biot-Savart considerations for atom chip wires, as used in our experiments, yield magnetic field gradients of about~$200\rm\,G/mm$ at~$\sim100\rm\,\mum$ from the chip, which is the starting distance for most of our experiments. Accurate control of this initial position, which is also crucial for the success of the experiments, is ensured by accurate control of chip wire currents and the homogeneous magnetic field referred to above. In addition, the straight atom chip wires have very low inductance, thereby enabling the generation of well-defined magnetic force pulses with currents that are typically tens of~$\rm\mus$ long. Such pulses are, in principle, able to induce momentum changes of hundreds of~$\hbar k$.\footnote
{We express the momentum transfer in units of~$\hbar k$, a reference momentum of a photon with~$\rm1\,\mum$ wavelength, in order to compare with atom interferometry based on optical beam splitters.}
Our earliest implementations of these experimental characteristics~\cite{Machluf2013t} were improved in subsequent apparatus upgrades~\cite{Margalit2018t}. 

Since the experiments proceed after turning off the magnetic trap, the observation time is limited by the time-of-flight~(TOF) of the falling atoms and the field-of-view of our absorption imaging detection system. The latter is limited to about~$4\rm\,mm$, corresponding to a maximum~TOF of about~$28\rm\,ms$. The optical detection system has a spatial resolution of about~$\rm5\,\mum$, an important consideration for measuring spatial interference patterns~(Sec.~\ref{sec:halfloop}). Further experimental details may be found in several recent~Ph.D. theses from our laboratory~\cite{Machluf2013t,Margalit2018t,Amit2020t}.

The Stern-Gerlach beam splitter~(SGBS), first implemented in~\cite{Machluf2013}, begins with an equal superposition of~$\one$ and~$\two$ as described above and depicted schematically in Fig.~\ref{fig:chip}. We then apply a magnetic field gradient~$\nabla{\bf|B|}$ for duration~$T_1$, which creates a state-dependent force ${\bf F}_{m_F}=m_Fg_F\mu_B\nabla{\bf|B|}$ on the atomic ensemble, where~$\mu_B$, $g_F$, and~$m_F$ denote the Bohr magneton, the Land\'e factor, and the projection of the angular momentum on the quantization axis, respectively. 

The magnetic potential created by the atom chip can be approximated as a sum of a linear part with characteristic force~${\bf F}$ and a quadratic part with characteristic frequency~$\omega$.  After this magnetic gradient splitting pulse, the new state of the atoms is given by $\psi_f=\frac{1}{\sqrt{2}}(|1,p_1\rangle+|2,p_2\rangle)$, where~${\bf p}_i={\bf F}_iT_1$ ($i=1,2$). This state represents a coherent superposition of two distinct momentum states, which are then allowed to separate spatially, thereby completing the operation of momentum and spatial splitting. 

As we discuss further in the following sections, the~SGBS can be extended as a tool for~SGI. We describe two main configurations: a ``half-loop'' configuration in which the separated wavepackets are allowed to propagate freely, expand and eventually overlap, producing spatial interference patterns analogous to a double-slit experiment, and a ``full-loop'' configuration in which the wavepackets are actively re-combined, analogous to a Mach-Zehnder interferometer.

By applying additional pulses with different timing, these methods have been used to demonstrate, to the best of our knowledge, the first Stern-Gerlach spatial fringe interferometer (Sec.~\ref{sec:halfloop},~\cite{Machluf2013,Margalit2019}), the first full-loop Stern-Gerlach interferometer (Sec.~\ref{sec:fullloop},~\cite{Margalit2018x,Amit2019}), and several applications that we will describe in Sec.~\ref{sec:applications}, including experiments to simulate the effect of proper time on quantum clock interference~\cite{Margalit2015,Zhou2018}.

{\parindent=0pt
\begin{figure} 
\begin{centering}
  \includegraphics[width=0.47\textwidth]{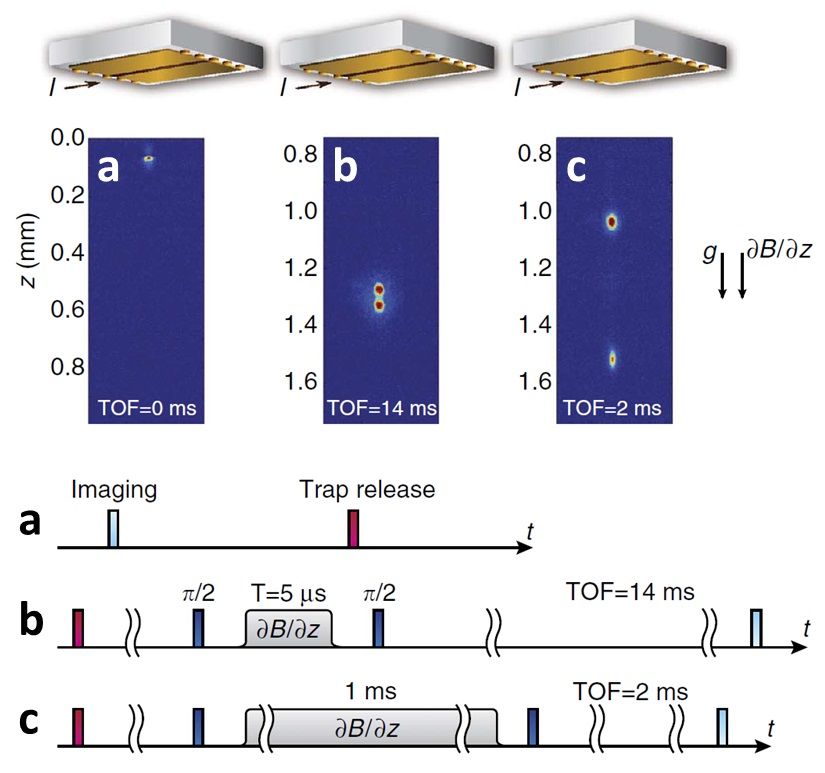}
	{\caption{\label{fig:chip} The Stern-Gerlach beam-splitter (SGBS) at work~\cite{Machluf2013,Machluf2013t}. SGBS~(a) input and~(b,c) output images, and the corresponding schematic descriptions. The top row depicts our atom chip, with a pulsed current~$I$ being used to generate the magnetic gradient~$\partial B/\partial z$ (we currently use three parallel wires with equal currents but opposing polarities). The chip faces downwards so that atoms can separate vertically during their free fall.  (a)~A magnetically trapped BEC in state~$\two$ before release. (b)~After a weak splitting of less than~$\hbar k$ using a~$\rm5\,\mus$ magnetic gradient pulse and allowing a~TOF of~$\rm14\,ms$. (c)~After a strong splitting of more than~$40\,\hbar k$ using a~$\rm1\,ms$ magnetic gradient pulse and allowing a~TOF of~$\rm2\,ms$. Interferometric signals are formed either as spatial interference fringes by passively allowing overlap of the wavepackets (the ``half-loop'' SGI), or as spin-state population oscillations upon actively recombining them (the ``full-loop'' SGI), as described in Secs.~\ref{sec:halfloop} and~\ref{sec:fullloop} respectively. Adapted from~\cite{Machluf2013}.}}
\end{centering}
\end{figure}
}

\negskip\negskip
\section{Half-loop Stern-Gerlach Interferometer}\negskip
\label{sec:halfloop}

The two separated wavepackets generated by the~SGBS initiate the pulse sequence shown in Fig.~\ref{fig:half}. Just after the~SG splitting pulse, another RF~$\pi/2$ pulse ($\rm10\,\mus$ duration) is applied, creating a wavefunction consisting of four wavepackets~\cite{Machluf2013t}, of which we are concerned only with the two~$\two$ wavepackets having momenta~${\bf p}_1$ and~${\bf p}_2$ [the~$\one$ components can be disregarded since they appear at different final positions on completing the pulse sequence and a time-of-flight~(TOF) period]. 

{\parindent=0pt
\begin{figure} 
\begin{centering}
   \includegraphics[width=0.47\textwidth]{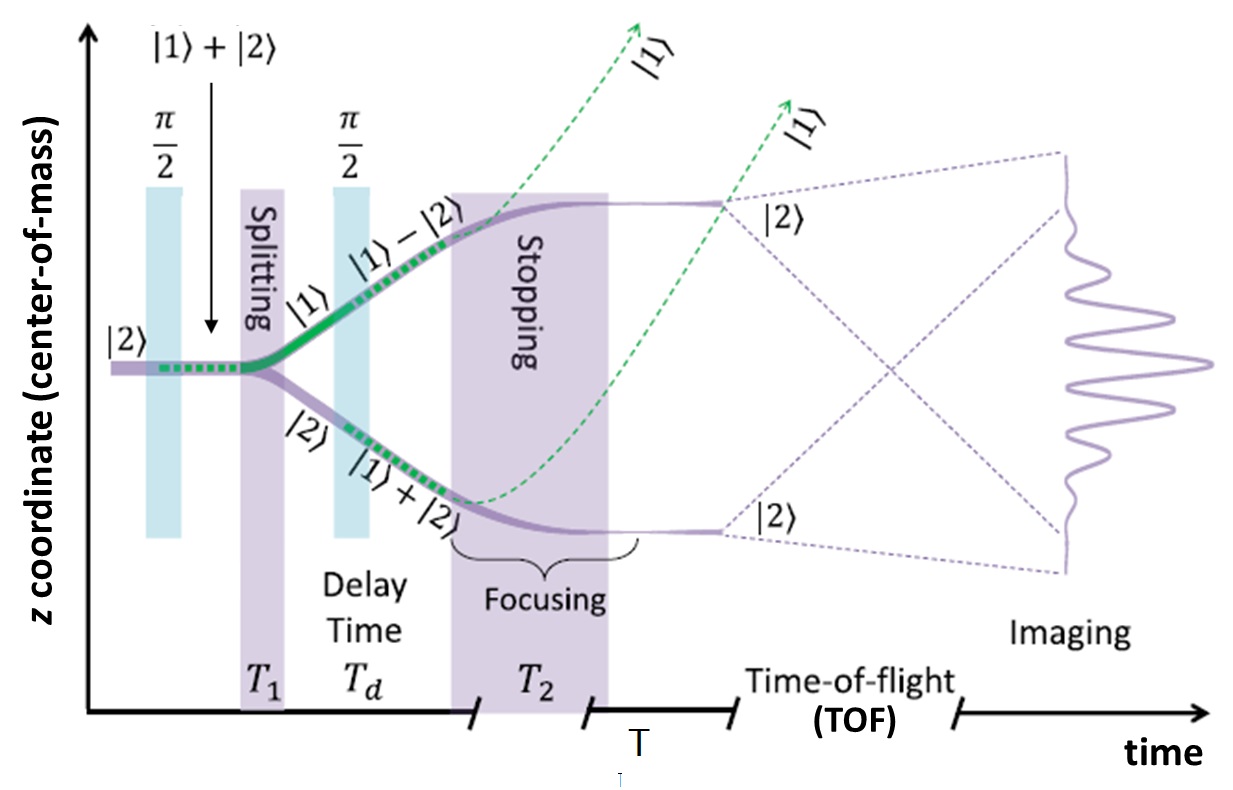} 
	{\caption{\label{fig:half} Schematic depiction of the longitudinal half-loop SGI giving rise to spatial interference fringes (vertical position~$z$ in the center-of-mass frame \vs\ time). The initial wavepacket~$\two$ (extreme left) is subjected to a~$\pi/2$ pulse (blue column) that transfers the atoms into the superposition state $\one+\two$. The first magnetic gradient pulse of duration~$T_1$ (purple column) induces a Stern-Gerlach splitting into~$\one$ (green curve) and~$\two$ (purple curve) having momenta~${\bf p}_1$ and~${\bf p}_2$, respectively. We then immediately apply a second~$\pi/2$ pulse that places these diverging~$\one$ and~$\two$ states into equal superpositions~$\one\mp\two$ as shown. The delay time~$T_d$ allows these wavepackets to spatially separate (in the~$z$ direction). The duration~$T_2$ of a second gradient pulse is tuned to bring the momentum difference between the~$\two$ components close to zero (see text), allowing their space-time trajectories to become parallel (solid purple curves) while expelling the~$\one$ components (dotted green trajectories). The atoms then fall freely under gravity. Given sufficient time-of-flight, the two~$\two$ wavepackets expand (dotted purple lines) and eventually overlap to generate spatial interference fringes, which are measured by taking an absorption image of the atoms. We note that due to the curvature of the magnetic field forming the magnetic gradient pulse, the long~$T_2$ pulse also focuses the wavepackets, as depicted in the figure. In fact, this focusing accelerates the process of final expansion, thereby creating the two-wavepacket overlap in a shorter time. Adapted from~\cite{Margalit2019} with permission \copyright~IOP Publishing~\& Deutsche Physikalische Gesellschaft. {\href{http://creativecommons.org/licenses/by/3.0/}{CC~BY~3.0}}}\negskip\negskip}  
\end{centering}
\end{figure}
}

The time interval between the two~RF pulses (in which there are only two wavepackets, each having a different spin) is reduced to a minimum ($\rm\sim40\,\mus$) to suppress the hindering effects of a noisy and uncontrolled magnetic environment, thereby removing the need for magnetic shielding.
After a magnetic gradient pulse of duration~$T_2$ designed to stop the relative motion of the two wavepackets, the atoms fall under gravity for a relatively long~TOF, expanding freely until they overlap to create spatial interference fringes as shown schematically in Fig.~\ref{fig:half} and experimentally in Fig.~\ref{fig:data}. 

The period of the interference fringes must be large enough to be observable with the spatial resolution of our imaging system (about~$\rm5\,\mum$). This is accomplished if two conditions are fulfilled. First, the distance between the two wavepackets,~$d$, should not be too large, since in principle the fringe periodicity varies as~$ht/md$ when the relative momentum is zero, where~$h$, $t$, and~$m$ are the Planck constant, TOF duration, and the atomic mass, respectively. Second, the momentum difference between the two wavepackets should be smaller than their momentum width to avoid orthogonality. This is accomplished by tuning the duration~$T_2$ of the second gradient pulse, which can stop the relative motion of the two~$\two$ wavepackets; despite being in the same spin state, the slower wavepacket experiences a stronger impulse than the faster one since it is considerably closer to the atom chip after the relatively long delay time~$T_d$. We have found that zeroing the momentum difference between the two wavepackets is very robust~\cite{Machluf2013t}.

Given that the final momentum difference between the two interfering wavepackets is smaller than their momentum spread, they overlap after a sufficiently long~TOF and an interference pattern appears with the approximate form:
\begin{multline}\label{eq:fit_function}
n(z,t) = A\exp\left[-\frac{(z-z_{\rm CM})^2}{2\sigma_z(t)^2}\right] \\[6pt] 
\times \left[1+V\cos\left(\frac{2\pi}{\lambda}(z-z_{\rm ref})+\phi\right)\right],
\end{multline}
where~$A$ is the amplitude, $z_{\rm CM}$ is the center-of-mass~(CM) position of the combined wavepacket at the time of imaging,~$\sigma_z(t)\approx\hbar t/2m\sigma_0$ is the final Gaussian width,~$\lambda\approx2\pi\hbar t/md$ is the fringe periodicity ($d=|z_1-z_2|$ is the distance between the wavepacket centers),~$V$ is the interference fringe visibility, and~$\phi=\phi_2-\phi_1$ is the global phase difference. The vertical position~$z$ is relative to a fixed reference point~$z_{\rm ref}$. The phases~$\phi_1$ and~$\phi_2$ are determined by an integral over the trajectories of the two wavepacket centers. We emphasize that Eq.~(\ref{eq:fit_function}) is not a phenomenological equation, but rather an outcome of our analytical model~\cite{Machluf2013}. 

In order to characterize the stability of the phase, which is the main figure of merit in interferometry, we average multiple experimental images with no post-selection or alignment (each single-shot image is a result of one experimental cycle). Large fluctuations in the phase and/or fringe  periodicity in a set of single-shot images would result in a low multi-shot visibility, while small fluctuations correspond to high multi-shot visibility. The multi-shot visibility is therefore a measure of the stability of the phase and periodicity. Single-shot and multi-shot visibilities are all extracted by fitting to Eq.~(\ref{eq:fit_function}) after averaging the experimental images along the~$x$ direction (see Fig.~\ref{fig:data}) to reduce noise. We note that these procedures have been used over several years of half-loop~SGI studies~\cite{Machluf2013,Margalit2019}, while the experimental results were simultaneously being greatly improved by significant modifications to the original apparatus~\cite{Machluf2013t,Margalit2018t}.

For a pure superposition state, as in our model, perfect fringe visibility~$V$ would be~1. A quantitative analysis of effects reducing~$V$ appears in~\cite{Margalit2018t,Margalit2019}. Some of these effects are purely technical, \eg\ imperfect~BEC purity and wavepacket overlap in~3D, as well as various imaging limitations~\etc\ Such technical effects are irrelevant to the phase and periodicity stability shown by the multi-shot visibility, so we normalize the latter to the mean of the single-shot visibilities taken from the same sample:~$V_N \equiv V_{\rm avg}/\langle V_s \rangle$, where~$V_{\rm avg}$ is the (un-normalized) visibility of the multi-shot average extracted from the fit, and~$\langle V_s \rangle$ is the mean visibility of the single-shot images which compose that multi-shot image. The normalized multi-shot visibility thus reflects shot-to-shot fluctuations of the global phase~$\phi$ and the fringe periodicity~$\lambda$. We note that some~BEC intrinsic effects, such as phase diffusion, would not lead to a reduction of the single-shot visibility, but may cause the randomization of the shot-to-shot phase. However, such effects are expected to be quite weak, since atom-atom interactions rapidly become negligible as the~BEC expands in free-fall, and the experiment may be described by single-atom
physics.

{\parindent=0pt
\begin{figure} 
\begin{centering}
   \includegraphics[width=0.47\textwidth]{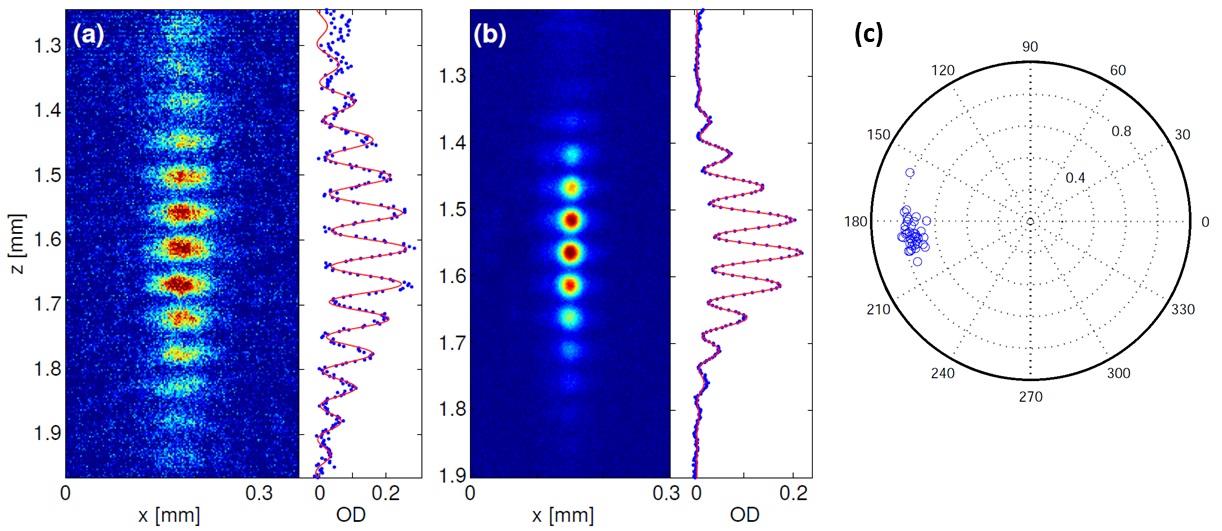} 
	{\caption{\label{fig:data} Spatial interference patterns from the Stern-Gerlach interferometer. (a) A single-shot interference pattern of a thermal cloud with a negligible~BEC fraction, fitted to Eq.~(\ref{eq:fit_function}) with a visibility of~$V=0.65$ (only slightly lower than single-shot visibilities typically measured for a BEC). (b) A multi-shot image made by averaging~40 consecutive interference images using a~BEC (no correction or post-selection) with a normalized visibility of~$V_N=0.99$. (c) Polar plot of phase~$0^\circ\le\phi\le360^\circ$ \vs\ visibility~$0\le V\le1$ obtained from fitting each of the~40 consecutive images averaged in~(b). The experimental parameters are $(T_1, T_d, T_2) =\rm(4, 116, 200)\,\mus$. Adapted from~\cite{Margalit2018t}.}\negskip}
\end{centering}
\end{figure}
}  

Representative results from the above analysis are shown in Fig.~\ref{fig:data}. The very high (normalized) visibility shown in~(b) demonstrates that the phase and periodicity are highly reproducible for each experimental cycle, the former being particularly emphasized in plot~(c). High-visibility fringes~($V>0.90$) were observed over a wide variety of experimental parameters, covering a range of maximum separations and velocities between the wavepackets. In particular, we conducted experiments at the apparatus-limited maximum value of~$T_d=\rm600\,\mus$ (which also required a long~TOF=$\rm21.45\,ms$) in order to maximize the spatial separation of the wavepackets during their time in the interferometer. These measurements achieved a separation~$d=\rm3.93\,\mum$, a factor of~20 larger than the atomic wavepacket size (after focusing, see Fig.~\ref{fig:half}), while maintaining a normalized visibility of~$V_N=0.90$~\cite{Margalit2019}. 

Given that our observed stable interference fringes arise from such well-separated paths, these experiments demonstrate what is, to the best of our knowledge, the first implementation of spatial~SG interferometry. This achievement is due to three main differences compared with previous~SG schemes. Firstly, we have used minimal-uncertainty wavepackets (a~BEC) rather than thermal beams. Secondly, while the splitting is based on two spin states, the wavepackets in the two interferometer arms are in the same spin state for most of the interferometric cycle, thus reducing their sensitivity to disruptive external magnetic fields. Finally, chip-scale temporal and spatial control allows the cancellation of path difference fluctuations. It should also be noted that a longitudinal~SGI, based on a particle beam source, cannot take images of spatial fringes due to the high velocity of the fringe pattern in the lab frame.

This, however, is not yet the four-field~SGI originally envisioned shortly after the original Stern-Gerlach experiments (as recounted in~\cite{Briegel1997}), since the separated wavepackets are not actively recombined in both position and momentum. The two remaining magnetic gradients required to complete such a ``closed''~SGI are discussed in the following section.  

{\parindent=0pt
\begin{figure*} 
\begin{centering}
   \includegraphics[width=0.67\textwidth]{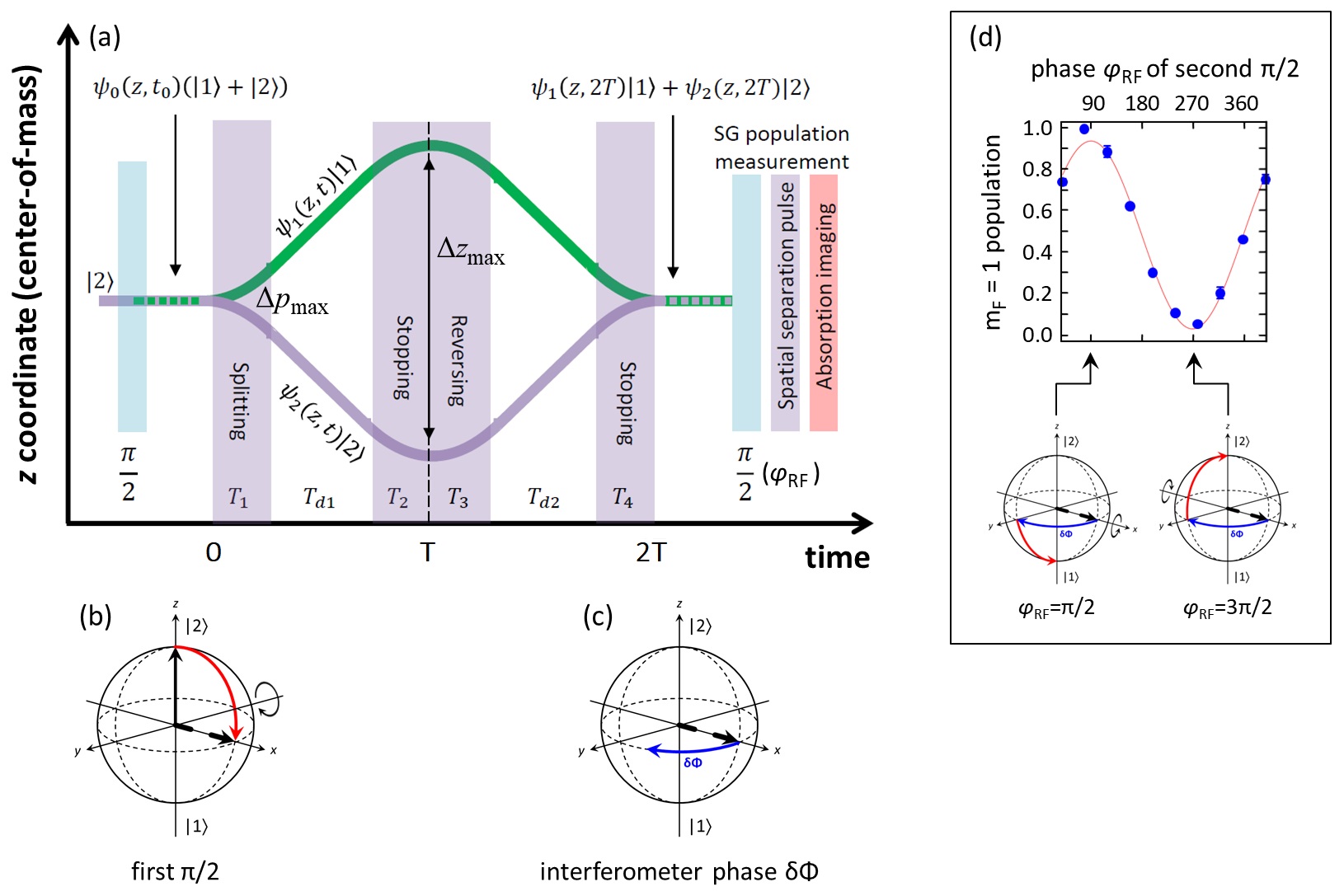} 
   \parbox{1.00\textwidth}{\caption{\label{fig:full} The longitudinal full-loop SGI giving rise to spin population oscillations, plotted in the center-of-mass frame as in Fig.~\ref{fig:half}. (a)~The sequence consists of~RF pulses (blue) to manipulate the inner (spin) degrees of freedom and magnetic gradients (purple) to control the momentum and position of the wavepackets. The interferometer is prepared from the initial wavepacket~$\two$ (extreme left) by applying a~$\pi/2$ pulse that transfers the atoms into the superposition state~$\one+\two$ [Bloch sphere shown in~(b)]. The first magnetic gradient pulse at~$t=0$ induces a Stern-Gerlach splitting into~$\one$ (green curve) and~$\two$ (purple curve). Three additional magnetic gradient pulses are used to stop the relative motion of the wavepackets (at their maximum separation~$\Delta z_{\rm max}$), reverse their momenta, and finally stop them at the same position along~$z$. The re-combined wavepacket at~$t=2T$ is therefore written as $\psi_1(z,2T)\one+\psi_2(z,2T)\two$, shown in~(c) for an arbitrary interferometer phase~$\delta\Phi$. After recombination, the population in~$\one$ is measured by applying a second~$\pi/2$ pulse with variable phase~$\varphi_{\rm RF}$, followed by a magnetic gradient to separate the populations and a subsequent pulse of the imaging laser. We expect to observe spin population fringes,~\ie\ oscillations in the~$m_F=1$ population, as we scan~$\varphi_{\rm RF}$, as indeed shown by the experimental results in~(d), for which the measured visibility is~95\%. The Bloch spheres in~(d) show the particular case in which the initial vector (dashed black arrow) acquires an interferometer phase~$\delta\Phi=\pi/2$ (blue arrow) followed by rotations about the~$+x$ ($\varphi_{\rm RF}=\pi/2$) or~$-x$ ($\varphi_{\rm RF}=3\pi/2$) axes respectively (red arrows). The states~$|F,m_F\rangle=|2,2\rangle\equiv\two$ and~$|2,1\rangle\equiv\one$ are defined along the~$z$ axis in the Bloch spheres. Adapted from~\cite{Margalit2018t}.
}}
\end{centering}
\end{figure*}
}

\negskip\negskip\negskip
\section{Full-loop Stern-Gerlach Interferometer}\negskip
\label{sec:fullloop}

Clearly, if a wavepacket can be coherently reconstructed after~SG splitting and recombination in a four-field configuration~\cite{Briegel1997}, it should be possible to observe an interference pattern at the output of such an~SGI. To the best of our knowledge however, no such interference pattern has heretofore been measured experimentally, and this is the task that we now describe, many details of which are taken from~\cite{Margalit2018t} and references therein. 

The device envisioned consists of four successive regions of magnetic gradients giving rise to the operations of splitting, stopping, reversing and, finally, stopping the two wavepackets, as shown schematically in Fig.~\ref{fig:full}(a). If executed perfectly, the two wavepackets would arrive at the output of such an interferometer with a minimal relative spatial displacement and momentum difference, so that an arbitrary initial spin state should be recoverable, using the spin state of the recombined wavepacket as the interference signal. However, the operation of such an interferometer was considered to be technically impractical, since coherent recombination of the two beam paths would require extremely precise control of the magnetic fields~\cite{Bohm1951}.

Our experiments begin, as before, with a~$\pi/2$ pulse creating a superposition of the two spin states~$\one$ and~$\two$ of~\Rb\ that is subsequently split into two momentum components by a magnetic gradient pulse (along the vertical axis~$z$) as described in Secs.~\ref{sec:SGBS} and~\ref{sec:halfloop}. Additional magnetic gradient pulses are needed to ``close'' the loop of such an interferometer, \ie\ to overlap the wavepackets spatially and with zero relative momentum. To stop the relative motion of the two wavepackets after the first pulse, and to accelerate them backwards, we reverse the current on the atom chip, causing the force applied by the magnetic field gradient to be in the opposite direction. Alternatively, we can apply a spin inversion procedure by using a~$\pi$ Rabi pulse that inverts the population between the two internal states, following which a magnetic gradient pulse will then apply the opposite differential momentum to the two wavepackets. We obtain the signal with the help of a second~$\pi/2$ pulse, followed by a spin population measurement. We measure the visibility by scanning the phase~$\varphi_{\rm RF}$ of this~$\pi/2$ pulse. 

Our full-loop interferometer is implemented with an experimental system in which care is taken to reduce a wide range of hindering effects relative to our earliest work~\cite{Machluf2013}. For example, a new atom chip was installed, utilizing a 3-wire configuration to produce a quadrupole magnetic field whose zero is at the precise height of the~BEC. This reduces phase fluctuations by exposing the wavepackets to a weaker magnetic field while still generating strong magnetic gradients. 

The practical difficulty encountered in re-assembling the original wavefunction was named the Humpty-Dumpty~(HD) effect~\cite{Englert1988,Schwinger1988,Scully1989}, implying that the initial wavepacket breaks under the~SG field and cannot be reunited, as noted in the brief historical perspective given in Sec.~\ref{sec:introduction}. Quantitatively, the spin coherence, which is measurable as the visibility~$V$ of the observed spin fringes, is expressed as~\cite{Schwinger1988}\vspace{-\topsep}{\vskip0.1\baselineskip}
\begin{equation}\label{eq:HD}
V = \exp\left\{-\frac{1}{2}\left[\left(\frac{\Delta z(2T)}{\sigma_z}\right)^2 + \left(\frac{\Delta p_z(2T)}{\sigma_p}\right)^2\right]\right\},
\end{equation}
where~$\Delta z(2T)$ and~$\Delta p_z(2T)$ denote the mismatch between the wavepackets in their final position and momentum respectively, after the interferometer duration~$2T$ [Fig.~\ref{fig:full}(a)], and~$\sigma_z$ and~$\sigma_p$ are the corresponding initial wavepacket widths. Equation~(\ref{eq:HD}) summarizes the main result of the~HD papers in relation to our experimental observable. We emphasize that this reduction in visibility has nothing to do with effects of decoherence due to some coupling with the environment. We also note that the above~HD calculation is done for a minimal-uncertainty wavepacket. For the general case, one can identify~$l_z=\hbar/\sigma_p$ and~$l_p=\hbar/\sigma_z$ as the relevant scales for coherence~\cite{Briegel1997,Schwinger1988}, where~$l_z$ and~$l_p$ are the spatial coherence length and the momentum coherence width, respectively.

Let us discuss the meaning of this equation. The quantities~$\sigma_z$ and~$\sigma_p$ characterize the initial atomic wavefunction, and are thus microscopic quantities. The quantities~$\Delta z$ and~$\Delta p_z$ describe the experimental imprecision in the final recombination. In a ``good''~SG experiment (\ie\ one which allows ``unmistakable'' splitting~\cite{Schwinger1988}) the maximum values of splitting in position and momentum should be much larger than their respective initial widths, meaning they should be macroscopic. On the other hand, according to Eq.~(\ref{eq:HD}), a nearly perfect maintenance of spin coherence~($V\simeq1$) requires both~$\Delta z\ll\sigma_z$ and~$\Delta p_z\ll\sigma_p$. Consequently, Eq.~(\ref{eq:HD}) tells us that we need to recombine macroscopic quantities with a microscopic level of precision. This is the challenge facing~SG interferometer experiments.

It is interesting to note that in the half-loop experiments, we found that~$\Delta p_z$ can be quite large (rendering the trajectories during the~TOF period in Fig.~\ref{fig:half} slightly non-parallel) without significantly reducing the measured spatial interference fringe visibility, so the stability of the half-loop experiments cannot be used to examine the~HD equation. This robustness of the half-loop may also be understood by considering the fact that the expansion of the wavepackets creates an enhanced local coherence length, since for every region of space the~$k$ vector variance becomes smaller as~TOF increases (see also~\cite{Miller2005,Romero-Isart2017}).

A practical full-loop~SG experiment must consider and address two effects. First, as noted above, the~HD effect requires accurate recombination, namely, small~$\Delta z$ and~$\Delta p_z$. These small values must be maintained for many experimental cycles, and thus a high level of stability in these values is also important. Achieving accurate recombination means that the overlap integral, calculated in Eq.~(\ref{eq:HD}), will have a significant non-zero value. Second, one must maintain a stable interferometer phase~$\delta\Phi$, so that it has the same value shot-to-shot. This requires that the coupling to external magnetic noise is kept to a minimum, either by shielding the experiment and stabilizing the electronics (\eg\ responsible for the homogeneous magnetic fields), or by conducting the experiment extremely quickly so that such environmental fluctuations do not have time to introduce significant phase noise. 

Our full-loop~SGI yields a visibility up to~95\% [Fig.~\ref{fig:full}(d)], proving that we are able to use the~SG effect to build a full-loop interferometer as originally envisioned almost a century ago. We note three differences between our realization and the scheme considered in the~HD papers: 1.~We use a~BEC, which is a minimum-uncertainty wavepacket, whereas the~HD papers considered atomic beam experiments with large uncertainties on the order of~$\sigma_z\sigma_p\simeq10^3$; 2.~We implement fast magnetic gradient pulses generated by running currents on the atom chip, in contrast to using constant gradients from permanent magnets that were considered in the original proposals; 3.~Our interferometer is a~1D longitudinal interferometer, while the originally envisioned~SGI was~2D, \ie\ it enclosed an area.

The full-loop experiments include a wide variety of optimizations and checks (see~\cite{Margalit2018t} for additional details). To make sure the spin superposition is not dephased due to some slowly varying gradients in our bias fields, we add~$\pi$ pulses giving rise to an echo sequence. To access a larger region of parameter space and to ensure the robustness of our results, we use several different configurations by, for example, implementing the reversing pulse~($T_3$) by inverting the sign of the atom chip currents \vs\ inverting the spins with the help of~$\pi$ pulses. We also utilize a variety of magnetic gradient magnitudes, and scan both the splitting gradient pulse duration~$T_1$ and the delay time between the pulses~$T_d$. All results are qualitatively the same. For weak splitting we observe high visibility~($\sim95\%$), while for a momentum splitting equivalent to~$\hbar k$ the visibility is still high~($\sim75\%$), indicating that the magnet precision enabled coherent spin-state recombination to a high degree.

Finally, we briefly compare our experiments to previous work in an elaborate series of~SGI experiments over a period of~15 years using metastable atomic beams~\cite{Robert1991,Miniatura1991,Miniatura1992a,Robert1992,Miniatura1992b,Chormaic1993,Baudon1999,Boustimi2000,Lesegno2003} and, more recently, thermal and ultra-cold alkali atoms~\cite{Marechal2000,Rubin2004}. A detailed discussion is given in~\cite{Margalit2018t}. While these longitudinal beam experiments did observe spin-population interference fringes, the experiments reviewed here are very different. Most importantly, an analogue of the full-loop configuration was never realized, as only splitting and stopping operations were applied (\ie\ there was no recombination) and wavepackets emerged from the interferometer with the same separation as the maximal separation achieved within (see Fig.~2 of~\cite{Chormaic1993} and footnote~[10] of~\cite{Marechal2000}). We have not found anywhere in the many papers published by this group (only some of which are referenced here) evidence of four operations being applied as required for a full-loop configuration, whether the experiment was with longitudinal or transverse gradients. In addition, no spatial interference fringes were observed, as the spatial modulation they observed was a signature of multiple parallel longitudinal interferometers, each having its own individual relative phase between its two wavepackets.

To conclude, we have shown that a full-loop may be realized. In addition, as previously shown in Heisenberg's argument, the momentum splitting is the figure of merit in determining the phase dispersion. In our experiment, coherence is observed up to a momentum splitting as high as~$\Delta p_z(T_1)/\sigma_p=60$. However, in contrast, the visibility is more sensitive to spatial splitting and we achieve~$\Delta z(T)/\sigma_z=4$, much lower than for the half-loop, where we achieved~$\Delta z/\sigma_z=18$. The splitting is coherent but its limits in terms of the~HD effect are yet to be explored quantitatively. Many mysteries remain to be solved, such as why is the observed reduction not symmetric in momentum and spatial splitting, in contrast to Eq.~\ref{eq:HD}. A simple answer, which is yet to be examined in detail, is the existence of some sort of spatial decoherence mechanism due to the environment. 

Having now described the~SG beam-splitter, the~SG half-loop, and the~SG full-loop, we show in the next section how these techniques may be used for different applications.

\negskip\negskip\negskip
\section{Applications}\negskip 
\label{sec:applications}

The pulse sequence in the half-loop experiments creates two spatially separated wavepackets in the state~$\two$ with zero relative momentum [left-most frame of Fig.~\ref{fig:clockdata}(a-c)]. We now take advantage of the long free-fall period in the experiment (labelled~TOF in Fig.~\ref{fig:half} \ie\ after the ``stopping pulse'') to further manipulate these wavepackets while they are allowed to expand and ultimately to overlap. The experiments are based on imposing a differential time evolution between the two wavepackets, which we measure as the interference patterns generated upon their recombination.

In particular, we create a ``clock'' state for each of the two wavepackets by first applying an~RF pulse that prepares the atoms in a superposition of two Zeeman sublevels~$\one$ and~$\two$ whose coefficients depend on the Bloch sphere angles~$\theta$ and~$\phi$. This superposition state is a two-level system evolving with a known period, as in the regular notion of an atomic clock. The~RF pulse (duration~$T_R$) controls the value of~$C=\sin\theta$, while a subsequent magnetic gradient pulse (duration~$T_G$) controls the value of~$D_I=\sin(\phi/2)$ by changing the relative ``tick'' rate~$\Delta\omega$ of the two clock wavepackets, as illustrated in Fig.~\ref{fig:clockdata}(a-c). The quantities~$C$ and~$D_I$ describe the clock preparation quality and the ideal distinguishability between the two clock interferometer arms respectively, which we will find quantitatively useful in our discussion of clock complementarity [see Eqs.~(\ref{eq:clockprep}) and~(\ref{eq:distinguishability}) below]. We note that, although the magnetic gradient pulse applies a different~SG force to each of the states within the clock, we have evaluated this effect for our experimental parameters and find that it is smaller than our experimental error bars ($\leq2\%$, Supplementary Materials of~\cite{Zhou2018}). 

{\parindent=0pt
\begin{figure} 
\begin{centering}
   \includegraphics[width=0.47\textwidth]{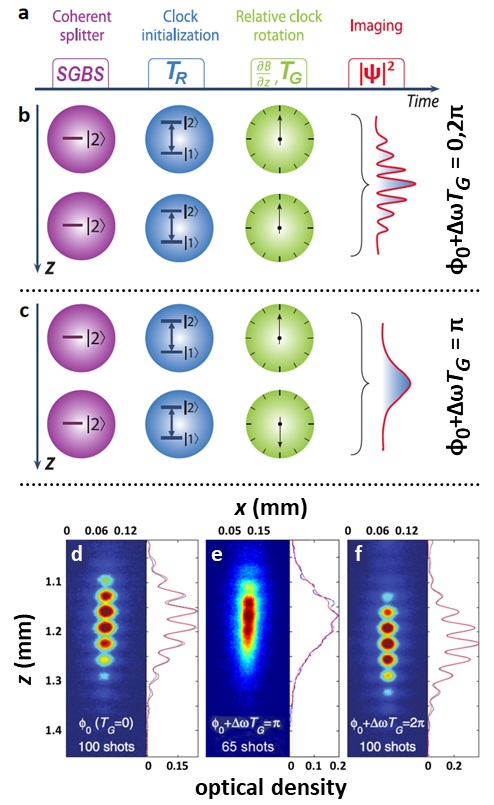} 
	{\caption{\label{fig:clockdata} Clock interferometry. (a)~Timing sequence (not to scale): Following a coherent spatial splitting by the~SGBS and a stopping pulse, the system consists of two wavepackets in the~$\two$ state (separated along the~$z$ axis) with zero relative velocity, as in Sec.~\ref{sec:halfloop}. The clock is then initialized with an~RF pulse of duration~$T_R$ (usually a~$\pi/2$ pulse, $T_R=\rm10\,\mus$) after which the relative ``tick'' rate~$\Delta\omega$ of the two clock wavepackets may be changed by applying a magnetic field gradient~$\partial B/\partial z$ of duration~$T_G$. Clock initialization occurs~$\rm1.5\,ms$ after trap release, the first~$\rm0.9\,ms$ of which is used for preparing the two wavepackets. The wavepackets are then allowed to expand and overlap and an image is taken. (b)~Evolution in time, synchronized with~(a). Each ball represents a clock wavepacket, where the hand represents its Bloch sphere phase~$\phi_{\rm BS}$. When the clock reading (\ie\ the position of the clock hand) in the two clock wavepackets is the same ($\phi_{\rm BS}=\phi_0+\Delta\omega T_G=0,2\pi$), fringe visibility is high. (c)~When the clock reading is opposite (orthogonal, $\phi_{\rm BS}=\phi_0+\Delta\omega T_G=\pi$), there is no interference pattern. (d)-(f)~Corresponding interference data of the two wavepackets \ie\ of the clock interfering with itself. All data samples are from consecutive measurements without any post-selection or post-correction. Single-shot patterns for~$\phi_{\rm BS}=\phi_0+\Delta\omega T_G=\pi$ also show very low fringe visibility (see Fig.~2(c) of~\cite{Margalit2015}). Adapted from~\cite{Margalit2015} and reprinted with permission from~AAAS; (e)~is adapted from~\cite{Margalit2018t}.}}  
\end{centering}
\end{figure}
}

\negskip\negskip\negskip
\subsection{Clock Interferometery}\label{subsec:clockinterference}\negskip

Let us first discuss the motivation for clock interferometry~\cite{Margalit2015}. Time in standard quantum mechanics~(QM) is a global parameter, which cannot differ between paths. Hence, in standard interferometry~\cite{Colella1975}, a height difference in a gravitational field between two paths would merely affect the relative phase of the clocks, shifting the interference pattern without degrading its visibility. In contrast, general relativity~(GR) predicts that a clock must ``tick'' slower along the lower path; thus if the paths of a clock passing through an interferometer have different heights, a time differential between the paths will yield ``which path'' information and degrade the visibility of the interference pattern according to the quantum complementarity relation between the interferometric visibility and the distinguishability of the wavepackets~\cite{Zych2011}. Consequently, whereas standard interferometry may probe~GR~\cite{Dimopoulos2007,Muntinga2013,Kuhn2014}, clock interferometry probes the interplay of~GR and~QM. For example, loss of visibility because of a proper time lag would be evidence that gravitational effects contribute to decoherence and the emergence of a classical world~\cite{Pikovski2015}.

Here we describe the use of this new tool~--~the clock interferometer~--~for its potential to investigate the role of time at the interface of~QM and~GR. Since the genuine~GR proper time difference is too small to be measured with existing experimental technology, our experiments instead simulate the proper time difference between the clock wavepackets using magnetic gradients, thereby causing the clock wavepackets to ``tick'' at different rates. Our results in this proof-of-principle experiment show that the visibility does indeed oscillate as a function of the simulated proper time lag. 

In the ultimate experiment, each part of the spatial superposition of a clock, located at different heights above Earth, would ``tick'' at different rates due to gravitational time dilation (so-called ``red-shift''). We can easily calculate the proper time difference between two arms of the clock interferometer as a figure-of-merit for  this effect. Using a first-order approximation of gravitational time dilation, and assuming a large separation between the arms of~$\Delta h=\rm1\,m$, an interferometer duration of~$T=\rm1\,s$ yields a proper time difference between the arms of only~$\Delta\tau \simeq Tg\Delta h/c^2 \simeq\rm10^{-16}\,s$. Such a small time difference means that a very accurate and fast-ticking clock must be sent through an interferometer with a large space-time area in order to observe the actual~GR effect. Both requirements are beyond our current experimental capabilities. Our ``synthetic'' red-shift is created by applying an additional magnetic gradient (of duration~$T_G$) that causes the clock wavepackets to ``tick'' at different rates. We denote the ``tick'' rate difference by~$\Delta\omega$.

Our results, some of which are presented in Fig.~\ref{fig:clockdata}(d-f), with more details in~\cite{Margalit2015,Margalit2018t}, show that the relative rotation between the two clock wavepackets affects the interferometric visibility. In the most extreme case, when the two clock states are orthogonal, \eg\ one in the state~$\frac{1}{\sqrt2}(\one+\two)$ and the other in the state~$\frac{1}{\sqrt2}(\one-\two)$, the visibility of the clock self-interference drops to near zero~[Fig.~\ref{fig:clockdata}(e)]. By varying the duration of the magnetic gradient~$T_G$ and thereby scanning the differential rotation angle~$\phi_{\rm BS}$ between the two clock wavepackets, we show quantitatively that the visibility oscillates as a function of our ``synthetic'' red-shift with a period of~$\Delta\omega T_G=2\pi$ [Fig.~\ref{fig:clockdata}(d,f)]. As an additional test of the clock interferometer, we modulate its preparation by changing the duration of the clock initialization pulse~$T_R$, which influences the relative populations of the two states composing the clock. This changes the state of the system from a no-clock state to a full-clock state in a continuous manner. The results show that the visibility behaves as expected in each case, further validating that it is the clock reading which is responsible for the oscillations in visibility that we observe as a function of~$T_R$~\cite{Margalit2015}.

\negskip\negskip\negskip
\subsection{Clock Complementarity}\label{subsec:clockcomplementarity}\negskip

These measurements of visibility may naturally be extended to study quantum complementarity for our self-interfering atomic clocks, which we again remark is at the interface of~QM and~GR. Our central consideration here is the inequality~\cite{Englert1996}\negskip\negskip
\begin{equation}\label{eq:complementarity}
V^2+D^2 \leq 1,
\end{equation}\negskip
where~$V$ is the ``visibility'' of an interference pattern such as discussed throughout this review, and~$D$ is the ``distinguishability'' of the two paths of the interfering particle. The latter quantity can also be measured directly in the clock experiments by controlling the angle~$\phi_{\rm BS}$, where~$(\theta=\pi/2,\phi_{\rm BS}=\Delta\omega T_G=\pi)$ prepares two perfectly distinguishable clocks such that~$D=1$ [Fig.~\ref{fig:clockdata}(e)]. A brief account of recent work theoretically and experimentally verifying this fundamental inequality is given by~\cite{Zhou2018} and references therein. 

It is important to investigate clock complementarity, particularly in view of recent theoretical work showing that spatial interferometers can be sensitive to a proper time lag between the paths~\cite{Giese2019} and speculation (see Table~1 in ~\cite{Zych2011}) that the inequality of Eq.~(\ref{eq:complementarity}) may be broken such that~$V^2+D^2>1$ when the effect of gravity is dominant. Zhou~\etal\ summarize the importance of this work as follows: ``\dots\ on the one hand, if the `ticking' rate of the clock depends on its path, then clock time provides which-path information and Eq.~(\ref{eq:complementarity}), developed in the framework of non-relativistic~QM, must apply. Yet, on the other hand, gravitational time lags do not arise in non-relativistic~QM, which is not covariant and therefore not consistent with the equivalence principle~\cite{Lugli2017t}. Hence our treatment of the clock superposition is a semiclassical extension of quantum mechanics to include gravitational red-shifts.''

The experiments we conducted in~\cite{Zhou2018} set out to test Eq.~(\ref{eq:complementarity}) quantitatively. Imperfect clock preparation (\ie\ with~$\theta\neq\pi/2$) reduces the measurable distinguishability~$D$ from its ideal value~$D_I$ as
\begin{flalign}
& D^2 = (C \cdot D_I)^2, \,\ \hbox{where}\,\   C    \equiv\sin\theta = 2\sqrt{P(1-P)} \label{eq:clockprep}          \\[6pt]
&	\!\!\!\!\!\hbox{and}\qquad\qquad\qquad       D_I  =\left|\sin(\Delta\phi/2)\right|  \label{eq:distinguishability}
\end{flalign}
with~$P$ and~$1-P$ denoting the populations (occupation probabilities) of the two energy eigenstates of the clock and~$\Delta\phi_{\rm BS}\equiv\phi_{\rm BS}^u-\phi_{\rm BS}^d$, where~$u$ and~$d$ denote the upper and lower paths of the interferometer, respectively. 

{\parindent=0pt
\begin{figure} 
\begin{centering}
   \includegraphics[width=0.47\textwidth]{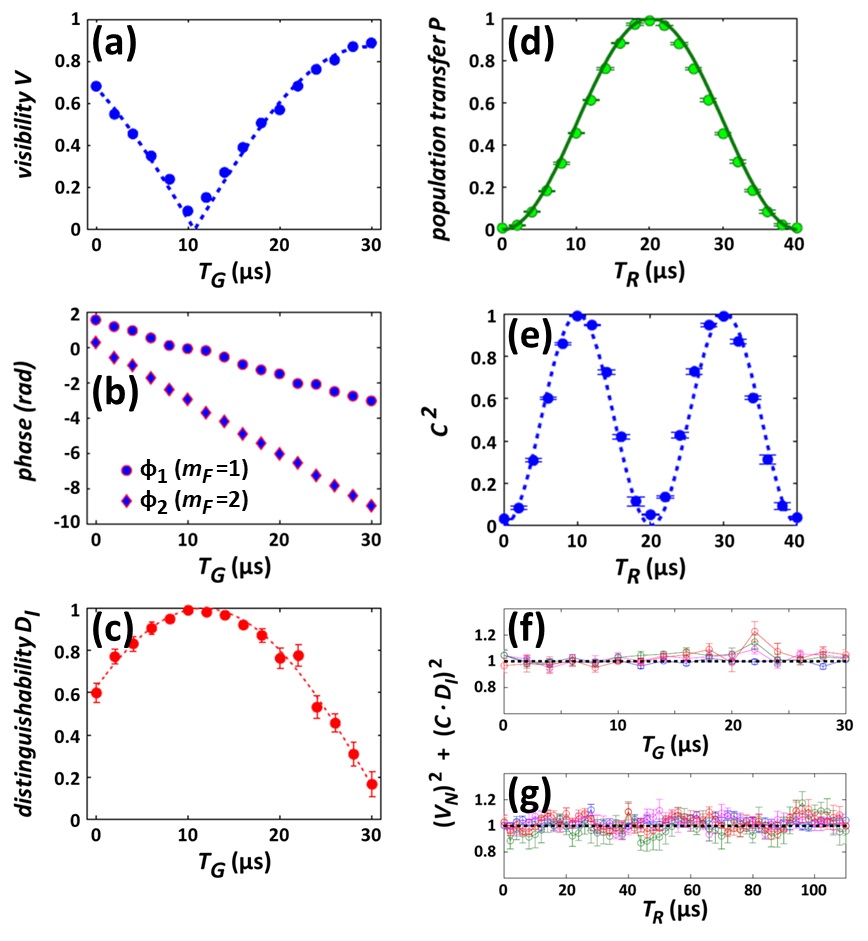} 
	{\caption{\label{fig:compdata} Clock complementarity: (a-e)~$V$, $D_I$, and~$C$ measured independently and (f-g)~combined in the complementarity relations of Eqs.~(\ref{eq:complementarity})-(\ref{eq:verified}). (a)~The visibility of an ideal clock~($C=1$) interference pattern \vs~$T_G$, fitted to~$\left|\cos(\phi/2)\right|$; (b-c)~the distinguishability is calculated from Eq.~(\ref{eq:distinguishability}) using the difference in relative angles~$\phi_2-\phi_1$, each measured separately and shown in~(b); and (d-e)~the clock preparation quality~$C$ is calculated from Eq.~(\ref{eq:clockprep}) using the data in~(d). Finally,~(f) shows the combination of all three parameters~$(V_N)^2+(C\cdot D_I)^2$ for four values of~$C$ when~$D_I$ is scanned and~(g) shows the same combination for four values of~$D_I$ when~$C$ is scanned. Only the data point in~(f) for~$T_G$ near~$\rm22\,\mus$ differs from unity, due to a relatively large experimental error in measuring the interferometric phase. These data therefore verify clock complementarity. Adapted from~\cite{Zhou2018} with permission \copyright~IOP Publishing~\& Deutsche Physikalische Gesellschaft, all rights reserved.}}
\end{centering}
\end{figure}
}

The experiment now has the task of measuring the three quantities~$V$, $D_I$ and~$C$ independently. We use the normalized visibility~$V_N$ as discussed in Sec.~\ref{sec:halfloop}. We evaluate~$D_I$ independently by measuring the relative phases in two single-state interferometers, one for each of the two clock states, and we measure~$C$, also independently, in a separate experiment by measuring~$P$ after the clock is initialized. Our results for these independently-measured quantities are shown in Fig.~\ref{fig:compdata}(a), (c), and~(e), where the results in~(c) and~(e) are based on analyzing the data in~(b) and~(d) respectively. We then combine these three quantities in the complementarity expression 
\begin{equation}\label{eq:verified}
(V_N)^2+(C\cdot D_I)^2 \leq 1,
\end{equation}\negskip
whereupon we see from Fig.~\ref{fig:compdata}(f-g) that the complementarity inequality [Eq.~(\ref{eq:complementarity})] is indeed upheld for the clock wavepackets superposed on two paths through our~SG interferometer. 

While the relation in Eq.~(\ref{eq:verified}) is specific to clock complementarity, it is unusual in linking non-relativistic quantum mechanics with general relativity. A direct test of this complementarity relation will come when~$D_I$ reflects the gravitational red-shift between two paths which traverse different heights. 

\negskip\negskip\negskip
\subsection{Geometric Phase}\label{subsec:geometric}\negskip

The geometric phase due to the evolution of the Hamiltonian is a central concept in quantum physics and may become advantageous for quantum technology. In noncyclic evolutions, a proposition relates the geometric phase to the area bounded by the phase-space trajectory and the shortest geodesic connecting its end points~\cite{Samuel1988,Bhandari1991,Dijk2010}. The experimental demonstration of this geodesic rule proposition in different systems is of great interest, especially due to its potential use in quantum technology. Here, we report a novel experimental confirmation of the geodesic rule for a noncyclic geometric phase by means of a spatial~SU(2) matter-wave interferometer, demonstrating, with high precision, the predicted phase sign change and~$\pi$ jumps. We show the connection between our results and the Pancharatnam phase~\cite{Pancharatnam1956}.

In the clock complementarity application just described, we scanned the third~RF pulse (duration~$T_R$) to vary the clock preparation parameter~$C=\sin\theta$. In our case, a~$\pi/2$ pulse typically corresponds to~$T_R=\rm10\,\mus$, so~$T_R<\rm10\,\mus$ places the Bloch vector in the northern hemisphere of the Bloch sphere with~$P_1<P_2$, while~$10<T_R<\rm30\,\mus$ places the Bloch vector in the southern hemisphere~($P_1>P_2$) \ie~the selected hemisphere is a periodic function of~$T_R$ such that an unequal superposition of~$\one$ and~$\two$ is created for each of the wavepackets unless~$\theta$ lies on the equator. After applying this~RF pulse (with some chosen duration~$T_R$), we adjust the phase difference between the two superpositions by applying the third magnetic gradient pulse of duration~$T_G$. This rotates the Bloch vectors along the latitude that was selected by the~RF pulse to points~$A$ and~$B$ in the northern hemisphere (or~$A',B'$ in the southern hemisphere) as shown in Fig.~\ref{fig:geomdata}(a), thereby affecting the phase difference~$\Delta\phi_{\rm BS}$, which we simply call~$\Delta\phi$ hereafter.  

{\parindent=0pt
\begin{figure} 
\begin{centering}
   \includegraphics[width=0.47\textwidth]{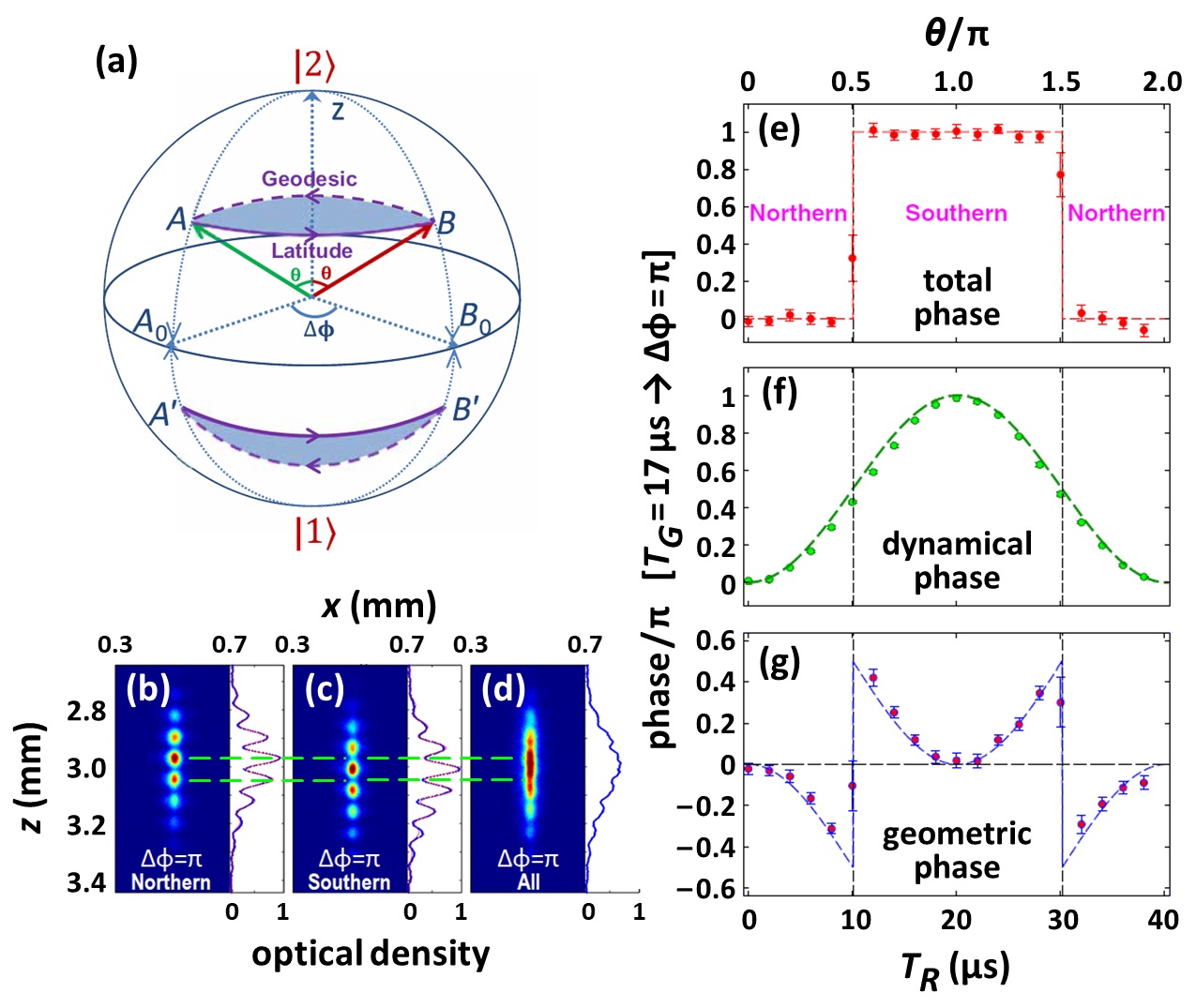} 
	{\caption{\label{fig:geomdata} Geometric phase. (a)~Bloch sphere for the two wavepackets (green and red arrows labeled~$A$ and~$B$, respectively) prepared by an~RF pulse (duration~$T_R$, rotation angle~$\theta$) and a subsequent magnetic gradient pulse (duration~$T_G$) that induces a rotation angle difference of~$\Delta\phi$. The rotation~$A\to B$ lies along a constant latitude (solid purple line), while the returning geodesic~$B\to A$ lies along the ``great circle'' curve (dashed purple line). Bloch vectors for corresponding wavepackets prepared in the southern hemisphere are shown as~$A'$ and~$B'$. (b-c)~Interference fringes generated by the half-loop~SGI, averaged over a total of~330 experimental shots with varying~$0<T_R<\rm40\,\mus$, while keeping a fixed value of~$T_G=\rm17\,\mus$ (this value of~$T_G$ corresponds to~$\Delta\phi=\pi$, see text). The dashed green lines show that the maxima in~(b) lie exactly where the minima occur in~(c), corresponding to Bloch vectors prepared in the northern and southern hemispheres, respectively. Adding all these interference patterns together in~(d) shows near-zero visibility, \ie\ they are completely out of phase. The fact that exactly the same pattern is observed while in the same hemisphere, independent of~$\theta$ (duration of~$T_R$), is called ``phase rigidity''. (e)~Total phase extracted from the interference fringes measured as a function of the~RF pulse duration (lower scale) and the corresponding latitude~$\theta$ (upper scale). Phase rigidity is clearly visible. (f-g)~Dynamical and geometric phases extracted from the data in~(e) and independently measured values of~$\theta$ and~$\Delta\phi$ (see text). The range of~$T_R$ in~(e-g) ($T_G$ is fixed at~$\rm17\,\mus$) corresponds to a full cycle from the northern hemisphere~($0<T_R<\rm10\,\mus$) through the southern hemisphere~($10<T_R<\rm30\,\mus$), and back to the north pole at~$T_R=\rm40\,\mus$. Adapted from~\cite{Zhou2020} with permission \copyright~the authors, some rights reserved; exclusive licensee AAAS. {\href{http://creativecommons.org/licenses/by/4.0/}{CC~BY~4.0}}}}
\end{centering} 
\end{figure}
}

The two wavepackets are allowed to interfere as in our half-loop experiments, enabling a direct measurement of the geometric phase. As usual, we extract the ``total'' interference phase (labeled~$\Phi$) by fitting the fringe patterns using Eq.~(\ref{eq:fit_function}). For general values of~$\theta$ and~$\Delta\phi$ (\ie\ after the application of both~$T_R$ and~$T_G$), we write the total phase between the two wavepackets as~\cite{Zhou2020}
\begin{equation}\label{eq:total}
\Phi=\arctan\left\{\frac{\sin^2(\theta/2)\,\sin(\Delta\phi)}{\cos^2(\theta/2)+\sin^2(\theta/2)\,\cos(\Delta\phi)}\right\}.
\end{equation}
Measurements of~$\Phi$, combined with values of~$\theta$ deduced independently from the relative populations of states~$\one$ and~$\two$, then allow us to fit~$\Delta\phi$ to high precision as a function of~$T_G$. These measurements verified that~$\Delta\phi$ depends linearly on~$T_G$, and we found that~$\Delta\phi=\pi$ occurs at~$T_G=\rm17\,\mus$. 

Figure~\ref{fig:geomdata}(b-c) shows interference fringe images for this specific value of~$T_G$, from which we extract the total phase as shown in Fig.~\ref{fig:geomdata}(e). We see immediately that this phase is independent of~$\theta$ within each hemisphere, an observation we call ``phase rigidity''. Moreover, the (constant) phase in each hemisphere differs by~$\pi$, which can also be deduced from the vanishing visibility shown in Fig.~\ref{fig:geomdata}(d) in which we have combined the data from both hemispheres. Evidently, there is a sharp jump in the phase of the interference pattern as~$\theta$ crosses the equator, as suggested by the singularities in Eq.~(\ref{eq:total}) that arise when~$\theta=\pi(n+1/2)$ (integer~$n$) and~$\Delta\phi=\pi$. 

To understand the non-cyclic geometric phase, we need to further examine the Bloch sphere. We see that the path from~$A\to B$ along the latitude~$\theta$ and returning along the geodesic (or ``great-circle route'') from~$B\to A$ encloses an area [blue shading in Fig.~\ref{fig:geomdata}(a)] in a counter-clockwise direction, whereas the corresponding path from~$A'\to B'$ and back again in the southern hemisphere proceeds in a clockwise direction. One-half of this area is the ``geometric phase'' that we now wish to calculate.

The total phase change~$\Phi$ for closed paths like~$A\to B\to A$ and~$A'\to B'\to A'$ is a sum of two contributions, the dynamical phase~$\Phi_D$ and the geometric phase~$\Phi_G$. The dynamical phase is given by~\cite{Samuel1988}
\begin{equation}\label{eq:dynamical}
\Phi_D=\frac{\Delta\phi}{2}\ \big(1-\cos\theta\big),
\end{equation}
which can be determined by measuring~$\theta$ and~$\Delta\phi$ independently. For the particular value of~$\Delta\phi=\pi$ chosen as a sub-set of our experimental data, we are then able to present~$\Phi_D$ in Fig.~\ref{fig:geomdata}(f). Finally, we subtract the phases~$\Phi_D$, as plotted in~(f), from the total phases~$\Phi$ plotted in~(e) (which, as noted above, are extracted directly from the observed interference pattern) to obtain the phases~$\Phi_G$. Namely, we perform~$\Phi-\Phi_D$ and get~$\Phi_G$, which is presented in Fig.~\ref{fig:geomdata}(g). Let us emphasize that the total phase~$\Phi$ is also the Pancharatnam phase~\cite{Pancharatnam1956}, and thus our experiment is also a direct measurement of this phase.

Our plot of~$\Phi_G$ exactly confirms the prediction shown in Fig.~4(d) of~\cite{Bhandari1991}, also reproduced as the dashed blue line in Fig.~\ref{fig:geomdata}(g). The predicted sign change as the latitude crosses the equator is clearly visible. The evident phase jump is due to the geodesic rule. When~$\Delta\phi=\pi$, the geodesic must go through the Bloch sphere pole for any~$\theta\neq\pi/2$. As the latitude approaches the equator (\ie\ increasing~$\theta$), the blue area in Fig.~\ref{fig:geomdata}(a) (twice~$\Phi_G$) grows continuously, reaching a maximum of~$\pi$ in the limit as~$\theta\to\pi/2$. As the latitude crosses the equator, the geodesic jumps from one pole to the other pole, resulting in an instantaneous change of sign of this large area and a phase jump of~$\pi$. 

Finally, our approach for testing the geodesic rule is unique for the following reasons: 1.~the use of a spatial interference pattern to determine the phase in a single experimental run (no need to scan any parameter to obtain the phase); 2.~the use of a common phase reference for both hemispheres while scanning~$\theta$, enabling verification of the~$\pi$ phase jump and the sign change; and 3.~obtaining the relative phase by allowing the two coherently-prepared wavepackets to expand in free flight and overlap, in contrast to previous atom interferometry studies that required additional manipulation of~$\theta$ and~$\Delta\phi$ to obtain interference. 

\negskip\negskip\negskip
\subsection{{\texorpdfstring{$T^3$}{}} Stern-Gerlach Interferometer}\label{subsec:T3SGI}\negskip

Here we consider an application of the full-loop~SGI wherein we minimize the delay times between successive~SG pulses as much as allowed by our electronics. In such an extreme scenario, it is expected that the phase accumulation will scale purely as~$T^3$, thus representing the first pure interferometric measurement of the Kennard phase~\cite{Amit2019} predicted in~1927~\cite{Kennard1927,Kennard1929} (see also~\cite{Rozenman2019,Rozenman2020,Rozenman2020x}). The theory for this experiment was done by the group of Wolfgang Schleich.

In order to describe the phase evolution of an atom moving in a time- and state-dependent linear potential, it is sufficient~\cite{Zimmermann2019} to know the two time-dependent forces~${\bf F}_u\equiv F_u(t){\bf e}_z$ and~${\bf F}_l\equiv F_l(t){\bf e}_z$ acting on the atom along the upper and lower branches, respectively, of the interferometer shown in Fig.~\ref{fig:Tcubed}, where~$z$ is the axis of gravity, the axis of our longitudinal interferometer, and also the axis of our magnetic gradients. 

In the present case, these forces comprise the gravitational force~$F_g=mg$ and the state-dependent magnetic forces ${\bf F}_i=-\mu_B(g_F)_i(m_F)_i\left(\partial{\bf|B|}/\partial z\right){\bf e}_z, (i=1,2)$:
\begin{equation}\label{eq:Ful}
  F_{u,l}(t) = F_g+F_{2,1}{\mathcal F}(t),
\end{equation}
where~$\mu_B$, $g_F$, and~$m_F$ are the Bohr magneton, the Land\'e factor, and the projection of the angular momentum on the quantization~($y$-)axis, respectively. The function~${\mathcal F}(t)$ provides the time-dependent modulation shown as the orange curve in Fig.~\ref{fig:Tcubed}(b):
\begin{multline}\label{eq:Ft}
{\mathcal F}(t)\equiv\Theta(t)-\Theta(t-T_1)-\Theta(t-T_1-T_d)+\Theta(t-3T_1-T_d) \qquad \\[6pt]
+\Theta(t-3T_1-2T_d)-\Theta(t-4T_1-2T_d).
\end{multline}
Here we are using the Heaviside step function~$\Theta(t)$ and we are assuming that the duration of each gradient pulse is identical, \ie\ $T_{2,3,4}=T_1$, as are the two delay times, $T_{d_1,d_2}=T_d$. We are also careful to ensure experimentally that the magnetic field is linear in the vicinity of the atoms and acts only along the vertical~($z$-)axis.\footnote
{Magnetic field linearity is ensured to a good approximation by the three-wire chip design and by carefully positioning the atoms very close to the center of the quadrupole field that they produce, as well as by the short distances that the atomic wavepackets travel~($\rm\sim1\,\mum$) compared to their distance from the chip~($\rm\sim100\,\mum$). We also adjust the duration of~$T_4$ slightly, relative to~$T_1$, to better optimize the visibility and account for any residual non-linearity. See~\cite{Amit2020t,Amit2019} for further details.}

{\parindent=0pt
\begin{figure} 
\begin{centering}
   \includegraphics[width=0.45\textwidth]{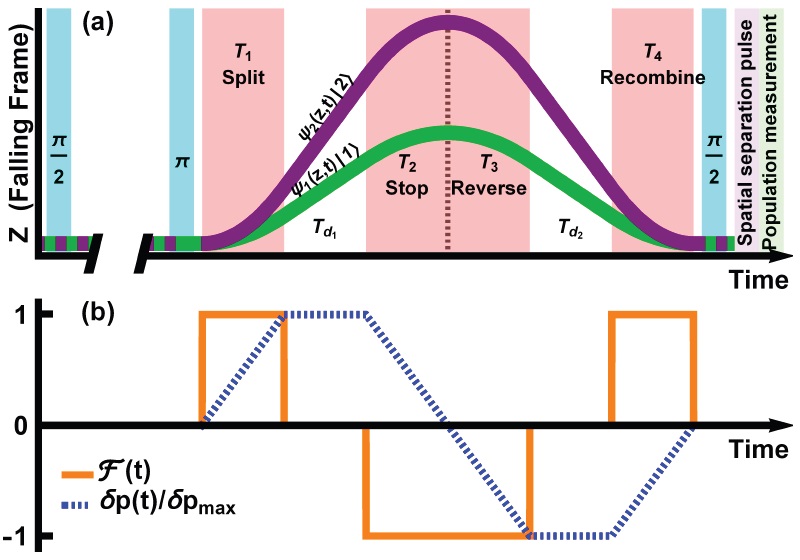} 
	{\caption{\label{fig:Tcubed} Pulse sequence of our longitudinal $T^3$-SGI (not~to scale). (a)~Trajectories of the atomic wavepackets with internal states~$\one$ (green curve) and~$\two$ (purple curve). Here we are using the freely-falling reference frame (gravity upwards), distinct from the center-of-mass reference frame used for Figs.~\ref{fig:half} and~\ref{fig:full}. Also shown are the~RF (blue) and magnetic gradient (red) pulses. The magnetic field gradients result in a state-dependent force along the $z$-direction while the strong bias magnetic field along the $y$-direction defines the quantization axis and ensures a two-level system. (b)~Time dependence of the relative force~${\mathcal F}={\mathcal F}(t)$ [orange curve, Eq.~(\ref{eq:Ft})] and the corresponding relative momentum~$\delta p(t)$ [blue dashed curve, Eq.~(\ref{eq:closedTcubed})] between the wavepackets moving along the two interferometer paths. In the experiment, we achieved the maximal separation~$\Delta z_{\rm{max}}=\rm1.2\,\mum$ in position and~$\Delta p_{\rm{max}}/m_{\rm{Rb}}=\rm17\,mm/s$ in velocity. Reprinted from~\cite{Amit2019} with permission \copyright~(2019) by the American Physical Society.}}
\end{centering}
\end{figure}
}

As in the full-loop~SGI experiments of Sec.~\ref{sec:fullloop}, we measure the spin population in state~$\one$ which, in this configuration, is a periodic function of the interferometer phase~\cite{Efremov2019x}.
\begin{equation}\label{eq:phiTcubed}
\begin{aligned}
              P_1         &= \frac{1}{2}\left[1-\cos\left(\delta\Phi+\varphi_0\right)\right], \qquad \\[6pt]
\hbox{where}\ \delta\Phi  &= \frac{1}{\hbar}\int_{0}^{T}dt\,\bar{F}(t)\delta z(t),
\end{aligned}
\end{equation}
with the total time~$T\equiv 4T_1+2T_d$. Note that the interferometer will be closed in both position and momentum provided that the differences 
\begin{equation}\label{eq:closedTcubed}
\begin{aligned}
            \delta p(t) &= \int_{0}^{t} d\tau \delta F(\tau)                    \\[6pt]
\hbox{and}\ \delta z(t) &= \frac{1}{m}\int_{0}^{t} d\tau \delta F(\tau)(t-\tau) \qquad
\end{aligned}
\end{equation}
both vanish at~$t=T$. Here~$\varphi_0$ is a constant phase taking into account possible technical misalignment, while~$\bar F(t)\equiv \left[F_u(t)+F_l(t)\right]/2 = F_g+\frac{1}{2}(F_1+F_2){\mathcal F}(t)$ and~$\delta F(t)\equiv F_u(t)-F_l(t) = (F_2-F_1){\mathcal F}(t)$ are the mean and relative forces respectively. From Eq.~(\ref{eq:phiTcubed}) we finally obtain 
\begin{multline}\label{eq:Tcubedphase}
\delta\Phi=\frac{mga_B}{\hbar}\left(\frac{\mu_1-\mu_2}{\mu_B}\right)\left(2T_1^3+3T_1^2T_d+T_1T_d^2\right) \\[6pt]
+\frac{ma_B^2}{\hbar}\left(\frac{\mu_1^2-\mu_2^2}{\mu_B^2}\right)\left(\frac{2}{3}T_1^3+T_1^2T_d\right),
\end{multline}
with~$a_B\equiv \mu_B\nabla B/m$ being the magnetic acceleration.

As sketched in Fig.~\ref{fig:Tcubed}, the experiment begins with an on-resonance~RF $\pi/2$-pulse that transfers the initially prepared internal atomic state~$\two$ to an equal superposition,~$\frac{1}{\sqrt2}(\one+\two)$. This~$\pi/2$ pulse is applied~$\rm1\,ms$ after the atoms are released from the trap in which they were prepared, in order to ensure that the trapping fields are fully quenched. Following a free-fall time of~$\rm400\,\mus$ (the first ``dark time''), we apply an~RF $\pi$-pulse that flips the atomic state to~$\frac{1}{\sqrt2}(\one-\two)$. After a second dark time of another~$\rm400\,\mus$, a second~$\pi/2$ pulse completes the spin-echo sequence. The~$\pi$-pulse inverts the population between the two states of the system thereby allowing any time-independent phase shift accumulated during the first dark time to be canceled in the second dark time. The experiment is completed by applying a magnetic gradient to separate the spin populations and a subsequent pulse of the detection laser to image both states simultaneously. 

As with all our previous full-loop experiments, the four magnetic field gradient pulses are produced by current-carrying wires on the atom chip. This magnetic pulse sequence sends the spin states~$\one$ and~$\two$ along different trajectories in the~SGI and ultimately closes the interferometer in both momentum and position. Careful calibration measurements verified that reversing the wire currents (the current flow is reversed during~$T_2$ and~$T_3$ relative to~$T_1$ and~$T_4$) provides magnetic accelerations that are equal in magnitude (but opposite in sign) to within our experimental uncertainty of~$<1\%$.

The experimental data shown in~Fig.~\ref{fig:Tcubedfit} are measured as a function of the time~$2<T_1<70\rm\,\mus$. From Eq.~(\ref{eq:Tcubedphase}), it is apparent that the~$T^3$ dependence will be most evident if~$T_d\ll T_1$, which is satisfied for most of the experimental range by using a fixed experimental value of~$T_d=\rm2.6\,\mus$ (limited by the speed of our electronic circuits). Note that~$T_1\lesssim\rm100\,\mus$ is limited by the duration of the second dark time. 

{\parindent=0pt
\begin{figure} 
\begin{centering}
   \includegraphics[width=0.47\textwidth]{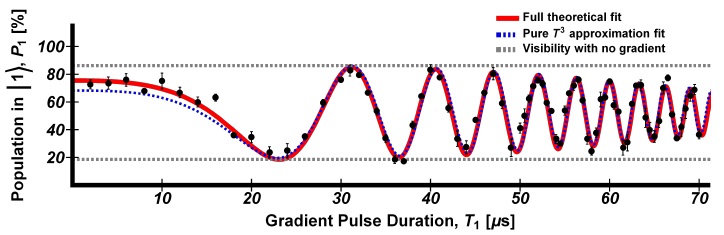} 
	{\caption{\label{fig:Tcubedfit} Measurement of the cubic phase with the $T^3$-SGI presented in Fig.~\ref{fig:Tcubed}. The solid red line represents a fit based on Eq.~(\ref{eq:Tcubedphase}), as described in the text. The dashed blue line is a fit with~$T_{d}=0$, showing that the interferometer phase scales purely as~$T_1^3$ for~$T_1\gtrsim\rm20\,\mus$. The visibility drops from~68\% to~32\% over~$\rm70\,\mus$ with a decay time of~$\rm75\,\mus$. This reduction results from inaccuracies in recombining the two interferometer paths. The dashed gray horizontal lines depict the maximal and minimal values of the population~$P_1$ measured independently without magnetic field gradients. Reprinted from~\cite{Amit2019} with permission  \copyright~(2019) by the American Physical Society.}}
\end{centering}
\end{figure}
}

The experimental data (dots) agree very well with the theory (solid red line) based on Eq.~(\ref{eq:Tcubedphase}), where the fitting parameters are the magnetic acceleration~$a_B$ as well as the decay constant of the visibility and a constant phase $\varphi_0$. The dashed blue line is obtained by setting~$T_{d}=0$, leading to a pure~$T_1^3$ scaling that is indistinguishable from the full theoretical fit for~$T_1\gtrsim\rm20\,\mus$:
\begin{equation}\label{eq:pureTcubed}
\delta\Phi^{(T^3)}\cong\frac{ma_B}{32\hbar}\left(\frac{\mu_1-\mu_2}{\mu_B}\right)\left(g+\frac{\mu_1+\mu_2}{3\mu_B}\,a_B\right)T^3.
\end{equation}

The maximum visibility displayed by the gray lines is first measured by performing only the~RF spin-echo sequence~($\pi/2-\pi-\pi/2$) without the magnetic field gradients and changing the phase of the second~$\pi/2$ pulse. The maximal visibility is limited by imperfections in the~RF pulses. As discussed above, utilizing an echo sequence allows us to cancel out contributions to the interferometer phase from the bias magnetic field, and to increase the coherence time. 

The excellent fit to these data allows a precise determination of the magnetic field acceleration, $a_{B}^{\rm fit}=\rm246.97\pm0.09\,m/s^2$. Separate measurements were used to independently determine the magnetic field gradient using time-of-flight~(TOF) techniques, which gave a value of~$a_{B}^{\rm TOF}=\rm249\pm2\,m/s^2$.\footnote
{These values for~$a_{B}^{\rm fit}$ and~$a_{B}^{\rm TOF}$ are different from those presented in~\cite{Amit2019} due to a  different fitting procedure used there. A full analysis and fitting procedures are presented in the Appendices of~\cite{Amit2020t}.}
While these measurements agree with one another, the difference in measurement errors clearly shows that our~$T^3$-SGI provides a much more precise measurement of the magnetic field gradient. 

Let us now consider the case when~$T_1\ll T_d$, such that during~$T_d$ the relative momentum~$\delta p_0\equiv ma_BT_1(\mu_1-\mu_2)/\mu_B$ between the paths is kept constant, \ie\ we take the magnetic field gradient pulses to be delta functions. 

In this limit the interferometer phase from Eq.~(\ref{eq:Tcubedphase}) becomes
\begin{equation}\label{eq:pureTsquared}
\delta\Phi^{(T^2)}\cong\frac{\delta p_0}{4\hbar}\,gT^2,
\end{equation}
scaling quadratically with the total time~$T\cong2T_d$, since we now maintain a piecewise constant momentum difference between the two arms. This is similar to the~$T^2$-SGI~\cite{Margalit2018x} or the Kasevich-Chu interferometer~\cite{Zimmermann2019}, although the momentum transfer~$\delta p_0$ is provided by the magnetic field gradient in the case of the~$T^2$-SGI, rather than by the laser light pulse. 

We conclude our discussion of this unique~$T^3$ interferometer by comparing the scaling of the interferometer phases~$\delta\Phi^{(T^3)}$ and~$\delta\Phi^{(T^2)}$ with the total interferometer time~$T$, as given by Eqs.~(\ref{eq:pureTcubed}) and~(\ref{eq:pureTsquared}) respectively. The data in Fig.~\ref{fig:scaling} are taken from Fig.~\ref{fig:Tcubedfit} and from our~$T^2$-SGI (when experimentally realizing the condition~$T_1 \ll T_d$), showing clearly that the~$T^3$-SGI significantly outperforms the~$T^2$-SGI with respect to total phase accumulation, even though the latter can currently operate for total times~$T$ up to three times larger than the former. Finally, let us briefly note that this~$T^3$ realization has already been coined a proof-of-principle experiment for testing the quantum nature of gravity~\cite{Marletto2020}.

{\parindent=0pt
\begin{figure} 
\begin{centering}
   \includegraphics[width=0.47\textwidth]{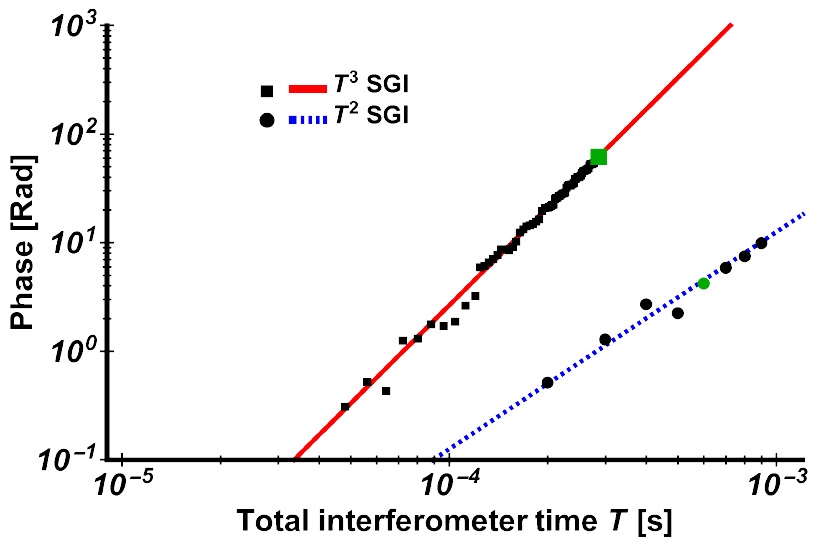} 
	{\caption{\label{fig:scaling} Scaling of the interferometer phases~$\delta\Phi^{(T^3)}$ [squares, Eq.(~\ref{eq:pureTcubed})] and~$\delta\Phi^{(T^2)}$ [circles, Eq.~(\ref{eq:pureTsquared})], as functions of the total interferometer time~$T$. The solid red line is fitted to our data for the~$T^3$-SGI and the dashed blue line is fitted to our~$T^2$-SGI data when experimentally realizing the condition~$T_1 \ll T_d$. In its current configuration with~$T_{\rm max}=\rm285\,\mus$, the phase of the~$T^3$-SGI is almost six times larger than the phase of the best~$T^2$-SGI, even though the magnetic field gradients and the maximal time~$T_{\rm max}=\rm924\,\mus$ are larger than those of the~$T^3$-SGI by factors of~2.3 and~3.2, respectively. For reference, the green square and green dot represent data for which the observed visibility is~$\approx\rm30\%$ for both the~$T^3$-SGI and~$T^2$-SGI respectively. Adapted from~\cite{Amit2019} with permission \copyright~(2019) by the American Physical Society.}} 
\end{centering}
\end{figure}
}

Looking into the future, we may ask if one may extend the~$T^3$ scaling to yet higher powers of time. In the Ramsey-Bord\'e interferometer~\cite{Borde1989}, the phase shift that scales linearly with the interferometer time~$T$ originates from a constant position difference between two paths during most of this time. In the Kasevich-Chu interferometer~\cite{Kasevich1991,Peters2001}, the quadratic scaling of the phase with time is caused by a piecewise constant velocity difference, while a piecewise constant acceleration difference between the two paths results in the cubic phase scaling~$\delta\Phi\propto T^3$, as presented above.

One can generalize this idea to achieve any arbitrary phase scaling by having a piecewise difference in the~$n^{\rm th}$ derivative of the position difference between the two paths. By designing an interferometer sequence consisting of pulses with a higher-order time-dependence of the forces, combined with careful choices of the relative signs and durations of the pulses, the total phase can be made to scale with the interferometer time as~$T^{n+1}$ for any chosen~$n>2$.

\negskip\negskip\negskip
\section{Outlook}\negskip 
\label{sec:outlook}

\subsection{SGI with Single Ions}\label{subsec:ionSGI}\negskip

The discovery of the Stern-Gerlach effect led to lively discussions early in the quantum era regarding the possibility of measuring an analogous effect for the electron itself (see \eg~\cite{Brillouin1928,Bohr1932}). The Lorentz force adds the complicating factor of a purely classical deflection of the electron beam that would smear out any expected~SG splitting. Here we summarize a generalized semiclassical discussion for any charged particle of mass~$m$ and charge~$e$ from~\cite{Henkel2019} (though with the co-ordinate system in Table~\ref{table:beams}). Assuming a beam momentum~$p_x$ and a transverse beam spatial width~$\Delta z$, we calculate the spread of the Lorentz force~$\Delta F_{\rm L}$ due to a transverse magnetic gradient~$B'$ as 
\begin{equation}
\Delta F_{\rm L} = \frac{e}{m}\ p_xB'\Delta z.
\end{equation}
Since the beam would be well collimated, $\Delta p_z<p_x$, so
\begin{equation}\label{eq:Lorentz}
\Delta F_{\rm L} > \frac{e}{m}\ B' \Delta p_z\Delta z \ge \frac{e\hbar}{2m}\ B' = \frac{m_e}{m}\ \frac{e\hbar}{2m_e}\ B' = \frac{m_e}{m}\ F_{\rm SG},
\end{equation}
where the second inequality uses the uncertainty principle and we have introduced the electron mass~$m_e$ to relate~$F_{\rm L}$ to the Stern-Gerlach force~$F_{\rm SG}$. 

The spatially inhomogeneous Lorentz broadening is therefore larger than the~SG splitting for electrons, at least in this semiclassical analysis~\cite{Gallup2001}, and this lively controversy has continued for decades though, as far as we know, without any conclusive experimental tests for electrons or for any other charged particles (see~\cite{Batelaan1997,Batelaan2002,Garraway2002} for reviews of the early history of this issue and recent perspectives). In contrast, Eq.~(\ref{eq:Lorentz}) shows no such fundamental problem if we take ions such that~$m_e/m<10^{-3}$, thereby motivating our proposals, including chip-based designs, for measurements using very high-resolution single ion-on-demand sources that have recently been developed using ultra-cold ion traps~\cite{Jacob2016,Jacob2016t}. As a practical matter, we note that a suitable ion chip could be fabricated and implemented either based on an array of current-carrying wires as analyzed in~\cite{Henkel2019} or on a magnetized microstructure like those implemented in~\cite{Hinds1999,Tran2019x}.

Although we did not extend our analysis to include the coherence of the spin-dependent splitting, the suggested ion-SG beam splitter may form a basic building block of free-space interferometric devices for charged particles. Here we quote from our collaborative work with Henkel, Schmidt-Kaler and co-workers~\cite{Henkel2019}. In addition to measuring the coherence of spin splitting as in the ``Humpty-Dumpty'' effect (see Sec.~\ref{sec:introduction}), we anticipate that such a device could provide new insights concerning the fundamental question of whether and where in the~SG device a spin measurement takes place. The ion interference would also be sensitive to Aharonov-Bohm phase shifts arising from the electromagnetic gauge field. The ion source would be a truly single-particle device~\cite{Jacob2016} and eliminate certain problems arising from particle interactions in high-density sources of neutral bosons~\cite{Tino2014b}. 

Such single-ion~SG devices would open the door for a wide spectrum of fundamental experiments, probing for example weak measurements and Bohmian trajectories. The strong electric interactions may also be used, for example, to entangle the single ion with a solid-state quantum device (an electron in a quantum dot or on a Coulomb island, or a qubit flux gate). This type of interferometer may lead to new sensing capabilities~\cite{Hasselbach2010}: one of the two ion wavepackets is expected to pass tens to hundreds of nanometers above a surface (in the chip configuration of our proposal~\cite{Henkel2019}) and may probe van der Waals and Casimir-Polder forces, as well as patch potentials. The latter are very important as they are believed to give rise to the anomalous heating observed in miniaturized ion traps~\cite{Brownnutt2015}. Due to the short distances between the ions and the surface, the device may also be able to sense the gravitational force on small scales~\cite{Behunin2014}. Finally, such a single-ion interferometer may enable searches for exotic physics. These include spontaneous collapse models, the fifth force from a nearby surface, the self-charge interaction between the two ion wavepackets, and so on. Eventually, one may be able to realize a double SG-splitter with different orientations, as originally attempted by Stern, Segr\`e and co-workers~\cite{Phipps1932,Frisch1933}, in order to test ideas like the Bohm-Bub non-local hidden variable theory~\cite{Bohm1966,Folman1994,Das2019}, or ideas on deterministic quantum mechanics (see, \eg~\cite{Schulman2017}). Since ions may form the basis of extremely accurate clocks, an ion-SG device would enable clock interferometry at a level sensitive to the Earth's gravitational red-shift (see the proof-of-principle experiments with neutral atoms in~\cite{Margalit2015,Zhou2018}). This has important implications for studying the interface between quantum mechanics and general relativity.

\negskip\negskip\negskip
\subsection{SGI with Massive Objects}\label{subsec:massiveSGI}\negskip

The main focus of our future efforts will be to realize an~SGI with massive objects. The idea of using the~SG interferometer, with a macroscopic object as a probe for gravity, has been detailed in several studies~\cite{Wan2016,Bose2017,Marshman2020a,Marshman2020b} describing a wide range of experiments from the detection of gravitational waves to tests of the quantum nature of gravity. Here we envision using a macroscopic body in the full-loop~SGI. We anticipate utilizing spin population oscillations as our interference observable rather than spatial fringes \ie\ density modulations. This observable, as demonstrated in the atomic~SGI described above, is advantageous because there is no requirement for long evolution times in order to allow the spatial fringes to develop, nor is high-resolution imaging needed to resolve the spatial fringes. Let us note that there are other proposals to realize a spatial superposition of macroscopic objects~\cite{Romero-Isart2017,Pino2018}.

As a specific example, let us consider a solid object comprising~$10^6\!-\!10^{10}$ atoms with a single spin embedded in the solid lattice, \eg\ a nano-diamond with a single~NV center. Let us first emphasize that even prior to any probing of gravity, a successful~SGI will already achieve at least~3 orders of magnitude more atoms than the state of the art in macroscopic-object interferometry~\cite{Fein2019}, thus contributing novel insight to the foundations of quantum mechanics. Another contribution to the foundations of quantum mechanics would be the ability to test continuous spontaneous localization~(CSL) models (\eg~\cite{Romero-Isart2011} and references therein). 

When probing gravity, the first contribution of such a massive-object~SGI would simply be to measure little~$g$. As the phase is accumulated linearly with the mass, a massive-object interferometer is expected to have much more sensitivity to~$g$ than atomic interferometers being used currently (assuming of course that all other features are comparable). This is also a method to verify that a massive-object superposition can be created~\cite{Scala2013,Wan2016,Toros2020x}. A second contribution would measure gravity at short distances, since the massive object may be brought close to a surface while in one of the~SGI paths, thus enabling probes of the fifth force. Once the~SG technology allows the use of large masses, a third contribution will be the testing of hypotheses concerning gravity self-interaction~\cite{Hatif2020x,Pino2018}, and once large-area interferometry is also enabled, a fourth contribution would be to detect gravitational waves~\cite{Marshman2020b}. Finally, placing two such~SGIs in parallel next to each other will enable probes of the quantum nature of gravity~\cite{Bose2017,Marletto2017}. Let us emphasize that, although high accelerations may be obtained with multiple spins, we intend to focus on the case of a macroscopic object with a single spin, since the observable of such a quantum-gravity experiment is entanglement, and averaging over many spins may wash out the signal.

To avoid the hindering consequences of the~HD effect, one must ensure that the experimental accuracy of the recombination, as discussed in Sec.~\ref{sec:fullloop}, will be better than the coherence length. Obviously it is very hard to achieve a large coherence length for a massive object, but recent experimental numbers and estimates seem to indicate that this is feasible. Another crucial problem is the coherence time. A massive object has a huge cross section for interacting with the environment (\eg\ background gas), but the extremely short interferometer times, as discussed in this review, seem to serve as a protective shield suppressing decoherence. We are now preparing a detailed account of these considerations~\cite{Margalit2020x}.

\halfskip\centerline{\bf Disclosure Statement}

The authors declare that they have no competing financial interests.

\centerline{\bf Acknowledgments}

We wish to warmly thank all the members~--~past and present~--~of the Atom Chip Group at Ben-Gurion University of the Negev, and the team of the~BGU nano-fabrication facility for designing and fabricating innovative high-quality chips for our laboratory and for others around the world. The work at BGU described in this review was funded in part by the Israel Science Foundation~(1381/13 and~1314/19), the~EC ``MatterWave'' consortium~(FP7-ICT-601180), and the German DFG through the DIP program (FO 703/2-1). We also acknowledge support from the~PBC program for outstanding postdoctoral researchers of the Israeli Council for Higher Education and from the Ministry of Immigrant Absorption (Israel).

\bibliography{OttoStern}

\end{document}